\documentclass[10pt,a4paper,reqno,oneside]{amsart}
\usepackage{tabls}
\usepackage{url}
\usepackage[linktocpage = true]{hyperref}
\hypersetup{
	colorlinks = true,
	citecolor = blue,
}
\usepackage{color}
\definecolor{red}{rgb}{1,0,0}
\definecolor{green}{rgb}{0,1,0}
\definecolor{blue}{rgb}{0,0,1}
\definecolor{refkey}{gray}{.625}
\definecolor{labelkey}{gray}{.625}
\usepackage[lite]{amsrefs}
\usepackage[a4paper]{geometry}
\usepackage{amsfonts}
\usepackage{amssymb}
\usepackage{mathrsfs}
\usepackage{amsmath}
\numberwithin{equation}{section}
\usepackage{amsthm}
\usepackage{amscd}
\usepackage{enumerate}
\usepackage{paralist}
\usepackage{xypic}
\usepackage{graphicx}
\usepackage{mathtools}
\usepackage[all,cmtip]{xy}
\usepackage{kantlipsum}
\usepackage{stmaryrd}
\usepackage{extarrows}
\usepackage{verbatim}
\allowdisplaybreaks

\makeatletter
\def\title@font{\normalsize\bfseries}
\let\ltx@maketitle\@maketitle
\def\@maketitle{\bgroup%
	\let\ltx@title\@title%
	\def\@\title{\resizebox{\textwidth}{!}{%
			\mbox{\title@font\ltx@title}%
	}}%
	\ltx@maketitle%
	\egroup}
\makeatother

\theoremstyle{plain}
\newtheorem*{zorn*}{Zorn's lemma}
\newtheorem*{tychonoff*}{Tychonoff's theorem}

\newtheorem*{theorem*}{Theorem}

\newtheorem{def-prop}[definition]{Definition-Proposition}

\newtheorem{prop-def}[definition]{Proposition-Definition}

\newcommand {\emptycomment}[1]{}

\begin{document}
	\title{The nonisospectral integrable hierarchies associated with Lie algebra $\mathfrak{sp}(6)$}
	
	\author{Yanhui Bi}
	\address{Center for Mathematical Sciences, College of Mathematics and Information Science,
		Nanchang Hangkong University}
	\email{biyanhui0523@163.com}
	\author{Bo Yuan}
	\address{College of Mathematics and Information Science,
		Nanchang Hangkong University}
	\email{yuanbo010806@163.com}
	\author{Yuqi Ruan}
	\address{College of Mathematics and Information Science,
		Nanchang Hangkong University}
	\email{ruanyuqi1023@163.com}
	\author{Tao Zhang}
	\address{College of Mathematics and Information Science,
		Henan Normal University}
	\email{zhangtao@htu.edu.cn}
	
	\begin{abstract}
		In this paper, we consider nonisospectral problems of two distinct dimensions on the loop algebra of the symplectic Lie algebra $\mathfrak{sp}(6)$, and construct two integrable systems. Furthermore, we derive their Hamiltonian structures using the Tu scheme. Additionally, we construct an integrable hierarchy on the generalized Lie algebra $\mathfrak{Gsp}(6)$ and establish its Hamiltonian structure as well.
	\end{abstract}
	\maketitle
	
	\section{Introduction}
	Since the 1960s, soliton theory and the study of nonlinear integrable systems have garnered significant attention from scientists worldwide. To this day, soliton theory and nonlinear integrable systems have found applications not only in branches of physics such as field theory, fluid mechanics, and nonlinear optics but also across various domains of natural sciences, including biology, chemistry, materials science, astrophysics, and communications. Up to now, researchers have discovered numerous intriguing properties in the $(1+1)$-dimensional and $(2+1)$-dimensional mathematical physics models under investigation. Building upon the analysis of loop algebras, Professor Gui-Zhang Tu proposed a systematic method--now known as the Tu Scheme--for deriving integrable hierarchies and their Hamiltonian structures starting from isospectral problems \cite{ref6}. This approach has led to the discovery of numerous significant integrable systems with profound mathematical and physical implications \cite{ref14,ref17}. Indeed, classical integrable systems--such as the Korteweg-de Vries (KdV) equation and the nonlinear Schr\"odinger (NLS) equation--typically arise from isospectral problems, where the spectral parameter remains invariant. However, in realistic physical scenarios, time-dependent parameters or spatially inhomogeneous environments often break this isospectrality, necessitating the study of nonisospectral integrable systems \cite{ref10,ref12,ref13,ref20}. In Ref \cite{ref1,ref2}, B.Y. He et al. constructed nonisospectral integrable systems on generalized Lie algebras and Lie superalgebras, respectively. In Ref \cite{ref9}, H.F. Wang and Y.F. Zhang introduced an efficient method for generating nonisospectral integrable hierarchies, and in Ref \cite{ref8}, they constructed two classes of nonisospectral integrable equation hierarchies and investigated their integrable couplings. In Ref \cite{ref11}, H.F. Wang and Y.F. Zhang construct a new class of higher-dimensional column-vector loop algebras and proposed a method for generating higher-dimensional isospectral and nonisospectral integrable hierarchies. In Ref \cite{ref3}, D. Levi extend the usual procedure for constructing hierarchies of integrable equations corresponding to nonisospectral deformations of a spectral problem to the case when the spectral parameter depends also on $x$. In Ref \cite{ref21}, Y.Z. Zhang et al. introduced  two kinds of appropriate isospectral problems by using a Lie algebra, and generated two new $(2+1)$-dimensional hierarchies of evolution equations. Starting from nonisospectral problems, we can likewise derive integrable equation hierarchies and their Hamiltonian structures \cite{ref1,ref16,ref19}. We first consider a nonisospectral problem
    \begin{align*}
    \begin{cases}
    \phi_{x}=U\phi,\\
    \phi_{t}=V\phi,\\
    \lambda_{t}=\sum_{i\geq 0}z_{i}(t)\lambda^{-i}
    \end{cases}
    \end{align*}
	From the compatibility condition of the Lax pair \cite{ref15}, one can derive the zero curvature representation $U_{t}-V_{x}+[U,V]=0$. Under the assumption that the elements of $U$(except for $\lambda$) are time-independent, solving the static zero-curvature equation $V_{x}=\frac{\partial U}{\partial\lambda}\lambda_{t}+[U,V]$ yields the solution for $V_{x}$. Proceeding further, we set $V_{+}^{n}=\sum_{0}^{n}(v_{i}^{1},\dots,v_{i}^{21})^{t}\lambda^{n-i}$. Solving the zero-curvature equation $\frac{\partial U}{\partial u}u_{t}+\frac{\partial U}{\partial\lambda}\lambda_{t}-V_{+x}^{n}+[U,V_{+}^{n}]$
	then yields the solution for $u_{t}$, where $v^{i}$ denotes an element of $V$, and $u=(u_{1},\dots,u_{12})^{t}$ denotes an element of $U$. The solution for $u_{t}$ constitutes an integrable hierarchy of equations. We proceed to construct the Hamiltonian structure for this integrable hierarchy using the trace identity \cite{ref6}
	\begin{align*}
		(\frac{\partial}{\partial u_{i}})\langle V,\frac{\partial U}{\partial \lambda}\rangle=(\lambda^{-\tau}(\frac{\partial}{\partial\lambda})\lambda^{\tau})\langle V,\frac{\partial U}{\partial u_{i}}\rangle,
	\end{align*}
	where $\langle x,y\rangle=tr(xy)$, $\tau=\frac{\lambda}{2}ln|tr(V^{2})|$. The systematic method for constructing integrable hierarchies and their Hamiltonian structures, as described above, is referred to as the Tu scheme. By selecting different matrix pairs $(U,V)$, one can generate a wide variety of distinct integrable hierarchies. In the pursuit of new integrable systems, it is customary to design spectral problems on loop algebras, a methodology that has yielded numerous significant integrable systems \cite{ref18}. In this work, we investigate nonisospectral problems defined on the loop algebra of the symplectic Lie algebra $\mathfrak{sp}(6)$. The $\mathfrak{sp}(6)$ is the Lie algebra comprising all $6\times 6$ complex matrices $X$ such that $X^{t}J+JX=0$, i.e.,
	\begin{align*}
		\mathfrak{sp}(6)=\{X\in\mathfrak{gl}(6,\mathbb{C})|X^{t}J+JX=0\},
	\end{align*}
	where is the standard symplectic form given by
	\begin{align*}
		J=\left(\begin{array}{cc}
			0&I_{3}\\
			-I_{3}&0
		\end{array}\right),
	\end{align*}
	and $I_{3}$ denotes the $3\times 3$ identity matrix. The loop algebra associated with the symplectic Lie algebra $\mathfrak{sp}(6)$ is constructed as follows:
	\begin{align*}
		\widetilde{\mathfrak{sp}(6)}=span\{E_{i}(n)\}_{i=1}^{21},
	\end{align*}
	where $\{E_{i}\}_{i=1}^{21}$ is a basis of $\mathfrak{sp}(6)$. In this work, we construct nonisospectral integrable systems by selecting distinct matrix pairs $(U,V)$ within the $\widetilde{\mathfrak{sp}(6)}$.
	\section{Two Nonisospectral Integrable Systems on the Lie Algebra $\mathfrak{sp}(6)$}\label{Sec:2}
	In this section, we construct two nonisospectral integrable systems on the loop algebra of the Lie algebra $\mathfrak{sp}(6)$, with dimensions (1+1) and (2+1), respectively.
	
	\subsection{Soliton Hierarchy Associated with $\mathfrak{sp}(6)$ in (1+1) dimension.\\}\label{Sec:2-1}
	The compact real form $\mathfrak{sp}(6)$ of complex symplectic Lie algebra $\mathfrak{sp}(6,\mathbb{C})$ is defined as
	\begin{align*}
		\mathfrak{sp}(6)=\{X\in\mathfrak{gl}(6,\mathbb{C})|X^{t}J+JX=0 \},
	\end{align*}
	where $J=\left(
	\begin{array}{cccc}
		0&I_{3}\\
		-I_{3}&0\\
	\end{array}\right)$, and $I_{3}$ is the $3\times3$ identity matrix. We can obtain the bases of Lie algebra $\mathfrak{sp}(6)$
	\begin{align*}
		&E_{1}=e_{11}-e_{44}, E_{2}=e_{22}-e_{55}, E_{3}=e_{33}-e_{66}, E_{4}=e_{12}-e_{54}, E_{5}=e_{21}-e_{45},\\
		& E_{6}=e_{13}-e_{64}, E_{7}=e_{31}-e_{46},E_{8}=e_{23}-e_{65}, E_{9}=e_{32}-e_{56}, E_{10}=e_{16}+e_{34},\\
		&  E_{11}=e_{43}+e_{61}, E_{12}=e_{25}, E_{13}=e_{52},E_{14}=e_{14},E_{15}=e_{41}, E_{16}=e_{15}+e_{24}, \\
		&E_{17}=e_{42}+e_{15}, E_{18}=e_{26}+e_{35}, E_{19}=e_{53}+e_{62}, E_{20}=e_{36}, E_{21}=e_{63},
	\end{align*}
	where $e_{ij}$ is a $6\times 6$ matrix with $1$ in the $(i, j)$-th position and $0$ elsewhere. The loop algebra of $\mathfrak{sp}(6)$ is defined as follows
	\begin{align*}
		\widetilde{\mathfrak{sp}(6)}=span\{E_{1}(n),\dots,E_{21}(n)\},
	\end{align*}
	where $E_{i}(n)=E_{i}\lambda^{n},\ [E_{i}(m),E_{j}(n)]=[E_{i},E_{j}]\lambda^{m+n}.$
	Let $U_{0},V_{0}\in\widetilde{\mathfrak{sp}(6)}$,
	\begin{align}
		U_{0}=\left(
		\begin{array}{cccccc}
			\lambda&0&0&u_{5}&u_{7}&u_{1}\\
			0&\lambda&0&u_{7}&u_{3}&u_{9}\\
			0&0&\lambda&u_{1}&u_{9}&u_{11}\\
			u_{6}&u_{8}&u_{2}&-\lambda&0&0\\
			u_{8}&u_{4}&u_{10}&0&-\lambda&0\\
			u_{2}&u_{10}&u_{12}&0&0&-\lambda
		\end{array}\right)\label{eq:2.1}
	\end{align}
	\begin{align}
		V_{0}=\left(
		\begin{array}{cccccc}
			a&d&f&p&r&k\\
			e&b&h&r&m&t\\
			g&j&c&k&t&v\\
			q&s&l&-a&-e&-g\\
			s&o&u&-d&-b&-j\\
			l&u&w&-f&-h&-c\\
		\end{array}\right)= \sum_{i\geq0}\left(
		\begin{array}{cccccc}
			a_{i}&d_{i}&f_{i}&p_{i}&r_{i}&k_{i}\\
			e_{i}&b_{i}&h_{i}&r_{i}&m_{i}&t_{i}\\
			g_{i}&j_{i}&c_{i}&k_{i}&t_{i}&v_{i}\\
			q_{i}&s_{i}&l_{i}&-a_{i}&-e_{i}&-g_{i}\\
			s_{i}&o_{i}&u_{i}&-d_{i}&-b_{i}&-j_{i}\\
			l_{i}&u_{i}&w_{i}&-f_{i}&-h_{i}&-c_{i}\\
		\end{array}\right)\lambda^{-i}\label{eq:2.2}
	\end{align}
	and consider a nonisospectral problem
	\begin{align*}
		\begin{cases}
			\phi_{x}=U_{0}\phi\\
			\phi_{t}=V_{0}\phi\\
			\lambda_{t}=\sum_{i\geq0}z_{i}(t)\lambda^{-i}.
		\end{cases}
	\end{align*}
	The stationary zero curvature representation $V_{0,x}=\frac{\partial U_{0}}{\partial\lambda}\lambda_{t}+[U_{0},V_{0}]$ gives
	\begin{align}
		\begin{cases}
			a_{x}=u_{1}l - u_{2}k + u_{5}q - u_{6}p  + u_{7}s - u_{8}r+z(t),\\
			b_{x}=u_{3}o - u_{4}m + u_{7}s- u_{8}r+ u_{9}u - u_{10}t+z(t) ,\\
			c_{x}=u_{1}l - u_{2}k + u_{9}u- u_{10}t +u_{11}w - u_{12}v+z(t)  ,\\
			d_{x}=u_{1}u- u_{4}r+ u_{5}s +u_{7}o - u_{8}p- u_{10}k ,\\
			e_{x}=- u_{2}t+ u_{3}s- u_{6}r + u_{7}q - u_{8}m +u_{9}l  ,\\
			f_{x}=u_{1}w- u_{2}p+u_{5}l+ u_{7}u-u_{10}r-u_{12}k	,\\
			g_{x}=u_{1}q-u_{2}v-u_{6}k-u_{8}t+u_{9}s+u_{11}l,\\
			h_{x}=- u_{2}r+u_{3}u+u_{7}l+u_{9}w-u_{10}m-u_{12}t,\\
			j_{x}=u_{1}s- u_{4}t- u_{8}k+  u_{9}o - u_{10}v + u_{11}u ,\\
			k_{x}=2\lambda k - u_{1}a- u_{1}c-u_{5}g - u_{7}j- u_{9}d - u_{11}f,\\
			l_{x}=- 2\lambda l+u_{2}a + u_{2}c+ u_{6}f + u_{8}h + u_{10}e + u_{12}g ,\\
			m_{x}=2\lambda m- 2u_{3}b - 2u_{7}e - 2u_{9}h ,\\
			o_{x}=- 2\lambda o+2u_{4}b + 2u_{8}d + 2u_{10}j ,\\
			p_{x}=2\lambda p - 2u_{1}f - 2u_{5}a - 2u_{7}d,\\
			q_{x}=- 2\lambda q+ 2u_{2}g+2u_{6}a + 2u_{8}e  ,\\
			r_{x}=2\lambda r -u_{1}h- u_{3}d - u_{5}e- u_{7}a- u_{7}b  - u_{9}f ,\\
			s_{x}=- 2\lambda s+ u_{2}j+u_{4}e+ u_{6}d+u_{8}a + u_{8}b  + u_{10}g  ,\\
			t_{x}=2\lambda t-u_{1}e-u_{3}j-u_{7}g-u_{9}b- u_{9}c - u_{11}h ,\\
			u_{x}=- 2\lambda u+ u_{2}d+u_{4}h+u_{8}f +u_{10}b + u_{10}c + u_{12}j ,\\
			v_{x}=2\lambda v-2u_{1}g-2u_{9}j-2u_{11}c,\\
			w_{x}=- 2\lambda w+ 2u_{2}f+2u_{10}h+2u_{12}c.\\
		\end{cases}\label{eq:2.3}
	\end{align}
	Take the initial values
	\begin{align*}
		a_{0}=\alpha(t), b_{0}=\beta(t), c_{0}=\gamma(t), d_{0}=e_{0}=\dots=v_{0}=w_{0}=z_{0}(t)=0,
	\end{align*}
	we have	
	\begin{align*}
		a_{1}=&b_{1}=c_{1}=z_{1}(t)x,\ d_{1}=\frac{1}{2}\partial^{-1}(u_{1}u_{10}+u_{4}u_{7}+u_{5}u_{8})(\beta(t)-\alpha(t)),\\
		e_{1}=&\frac{1}{2}\partial^{-1}(u_{2}u_{9}+u_{3}u_{8}+u_{6}u_{7})(\alpha(t)-\beta(t)),\ f_{1}=\frac{1}{2}\partial^{-1}(u_{1}u_{12}+u_{2}u_{5}+u_{7}u_{10})(\gamma(t)-\alpha(t)),\\
		g_{1}=&\frac{1}{2}\partial^{-1}(u_{1}u_{6}+u_{2}u_{11}+u_{8}u_{9})(\alpha(t)-\gamma(t)),\
		h_{1}=\frac{1}{2}\partial^{-1}(u_{2}u_{7}+u_{3}u_{10}+u_{9}u_{12})(\gamma(t)-\beta(t)),\\
		j_{1}=&\frac{1}{2}\partial^{-1}(u_{1}u_{8}+u_{4}u_{9}+u_{10}u_{11})(\beta(t)-\gamma(t)),\
		k_{1}=\frac{1}{2}u_{1}(\alpha(t)+\gamma(t)),\ l_{1}=\frac{1}{2}u_{2}(\alpha(t)+\gamma(t)),\\
		m_{1}=&u_{3}\beta(t),\ o_{1}=u_{4}\beta(t),\ p_{1}=u_{5}\alpha(t),\ q_{1}=u_{6}\alpha(t),\ r_{1}=\frac{1}{2}u_{7}(\alpha(t)+\beta(t)),\ s_{1}=\frac{1}{2}u_{8}(\alpha(t)+\beta(t)),\\
		t_{1}=&\frac{1}{2}u_{9}(\beta(t)+\gamma(t)),\ u_{1}=\frac{1}{2}u_{10}(\beta(t)+\gamma(t)),\
		v_{1}=u_{11}\gamma(t),\ w_{1}=u_{12}\gamma(t),\\
		k_{2}=&\frac{1}{4}u_{1x}(\alpha(t)+\gamma(t))+\frac{1}{4}u_{5}\partial^{-1}(u_{1}u_{6}+u_{2}u_{11}+u_{8}u_{9})(\alpha(t)-\gamma(t))\\
		&+\frac{1}{4}u_{7}\partial^{-1}(u_{1}u_{8}+u_{4}u_{9}+u_{10}u_{11})(\beta(t)-\gamma(t))+\frac{1}{4}u_{9}\partial^{-1}(u_{1}u_{10}+u_{4}u_{7}+u_{5}u_{8})(\beta(t)-\alpha(t))\\
		&+\frac{1}{4}u_{11}\partial^{-1}(u_{1}u_{12}+u_{2}u_{5}+u_{7}u_{10})(\gamma(t)-\alpha(t))+u_{1}z_{1}(t)x,\\
		l_{2}=&-\frac{1}{4}u_{2x}(\alpha(t)+\gamma(t))+\frac{1}{4}u_{6}\partial^{-1}(u_{1}u_{12}+u_{2}u_{5}+u_{7}u_{10})(\gamma(t)-\alpha(t))\\
		&+\frac{1}{4}u_{8}\partial^{-1}(u_{2}u_{7}+u_{3}u_{10}+u_{9}u_{12})(\gamma(t)-\beta(t))+\frac{1}{4}u_{10}\partial^{-1}(u_{2}u_{9}+u_{3}u_{8}+u_{6}u_{7})(\alpha(t)-\beta(t))\\
		&+\frac{1}{4}u_{12}\partial^{-1}(u_{1}u_{6}+u_{2}u_{11}+u_{8}u_{9})(\alpha(t)-\gamma(t))+u_{2}z_{1}(t)x,\\
		m_{2}=&\frac{1}{2}u_{3x}\beta(t)+\frac{1}{2}u_{7}\partial^{-1}(u_{2}u_{9}+u_{3}u_{8}+u_{6}u_{7})(\alpha(t)-\beta(t))\\
		&+\frac{1}{2}u_{9}\partial^{-1}(u_{2}u_{7}+u_{3}u_{10}+u_{9}u_{12})(\gamma(t)-\beta(t))+u_{3}z_{1}(t)x,\\
		o_{2}=&-\frac{1}{2}u_{4x}\beta(t)+\frac{1}{2}u_{8}\partial^{-1}(u_{1}u_{10}+u_{4}u_{7}+u_{5}u_{8})(\beta(t)-\alpha(t))\\
		&+\frac{1}{2}u_{10}\partial^{-1}(u_{1}u_{8}+u_{4}u_{9}+u_{10}u_{11})(\beta(t)-\gamma(t))+u_{4}z_{1}(t)x,\\
		p_{2}=&\frac{1}{2}u_{5x}\alpha(t)+\frac{1}{2}u_{1}\partial^{-1}(u_{1}u_{12}+u_{2}u_{5}+u_{7}u_{10})(\gamma(t)-\alpha(t))\\
		&+\frac{1}{2}u_{7}\partial^{-1}(u_{1}u_{10}+u_{4}u_{7}+u_{5}u_{8})(\beta(t)-\alpha(t))+u_{5}z_{1}(t)x,\\
		q_{2}=&-\frac{1}{2}u_{6x}\alpha(t)+\frac{1}{2}u_{2}\partial^{-1}(u_{1}u_{6}+u_{2}u_{11}+u_{8}u_{9})(\alpha(t)-\gamma(t))+u_{6}z_{1}(t)x\\
		&+\frac{1}{2}u_{8}\partial^{-1}(u_{2}u_{9}+u_{3}u_{8}+u_{6}u_{7})(\alpha(t)-\beta(t)),\\
		r_{2}=&\frac{1}{4}u_{7x}(\alpha(t)+\beta(t))+\frac{1}{4}u_{1}\partial^{-1}(u_{2}u_{7}+u_{3}u_{10}+u_{9}u_{12})(\gamma(t)-\beta(t))\\
		&+\frac{1}{4}u_{3}\partial^{-1}(u_{1}u_{10}+u_{4}u_{7}+u_{5}u_{8})(\beta(t)-\alpha(t))+\frac{1}{4}u_{5}\partial^{-1}(u_{2}u_{9}+u_{3}u_{8}+u_{6}u_{7})(\alpha(t)-\beta(t))\\
		&+\frac{1}{4}u_{9}\partial^{-1}(u_{1}u_{12}+u_{2}u_{5}+u_{7}u_{10})(\gamma(t)-\alpha(t))+u_{7}z_{1}(t)x,\\
		s_{2}=&-\frac{1}{4}u_{8x}(\alpha(t)+\beta(t))+\frac{1}{4}u_{2}\partial^{-1}(u_{1}u_{8}+u_{4}u_{9}+u_{10}u_{11})(\beta(t)-\gamma(t))\\
		&+\frac{1}{4}u_{4}\partial^{-1}(u_{2}u_{9}+u_{3}u_{8}+u_{6}u_{7})(\alpha(t)-\beta(t))+\frac{1}{4}u_{6}\partial^{-1}(u_{1}u_{10}+u_{4}u_{7}+u_{5}u_{8})(\beta(t)-\alpha(t))\\
		&+\frac{1}{4}u_{10}\partial^{-1}(u_{1}u_{6}+u_{2}u_{11}+u_{8}u_{9})(\alpha(t)-\gamma(t))+u_{8}z_{1}(t)x,\\
		t_{2}=&\frac{1}{4}u_{9x}(\beta(t)+\gamma(t))+\frac{1}{4}u_{1}\partial^{-1}(u_{2}u_{9}+u_{3}u_{8}+u_{6}u_{7})(\alpha(t)-\beta(t))\\
		&+\frac{1}{4}u_{3}\partial^{-1}(u_{1}u_{8}+u_{4}u_{9}+u_{10}u_{11})(\beta(t)-\gamma(t))+\frac{1}{4}u_{7}\partial^{-1}(u_{1}u_{6}+u_{2}u_{11}+u_{8}u_{9})(\alpha(t)-\gamma(t))\\
		&+\frac{1}{4}u_{11}\partial^{-1}(u_{2}u_{7}+u_{3}u_{10}+u_{9}u_{12})(\gamma(t)-\beta(t))+u_{9}z_{1}(t)x,\\
		u_{2}=&-\frac{1}{4}u_{10x}(\beta(t)+\gamma(t))+\frac{1}{4}u_{2}\partial^{-1}(u_{1}u_{10}+u_{4}u_{7}+u_{5}u_{8})(\beta(t)-\alpha(t))\\
		&+\frac{1}{4}u_{4}\partial^{-1}(u_{2}u_{7}+u_{3}u_{10}+u_{9}u_{12})(\gamma(t)-\beta(t))+\frac{1}{4}u_{8}\partial^{-1}(u_{1}u_{12}+u_{2}u_{5}+u_{7}u_{10})(\gamma(t)-\alpha(t))\\
		&+\frac{1}{4}u_{12}\partial^{-1}(u_{1}u_{8}+u_{4}u_{9}+u_{10}u_{11})(\beta(t)-\gamma(t))+u_{10}z_{1}(t)x,\\
		v_{2}=&\frac{1}{2}u_{11x}\gamma(t)+\frac{1}{2}u_{1}\partial^{-1}(u_{1}u_{6}+u_{2}u_{11}+u_{8}u_{9})(\alpha(t)-\gamma(t))\\
		&+\frac{1}{2}u_{9}\partial^{-1}(u_{1}u_{8}+u_{4}u_{9}+u_{10}u_{11})(\beta(t)-\gamma(t))+u_{11}z_{1}(t)x,\\
		w_{2}=&-\frac{1}{2}u_{12x}\gamma(t)+\frac{1}{2}u_{2}\partial^{-1}(u_{1}u_{12}+u_{2}u_{5}+u_{7}u_{10})(\gamma(t)-\alpha(t))\\
		&+\frac{1}{2}u_{10}\partial^{-1}(u_{2}u_{7}+u_{3}u_{10}+u_{9}u_{12})(\gamma(t)-\beta(t))+u_{12}z_{1}(t)x,\\
		\dots&\dots
	\end{align*}
	Now, taking
	\begin{align*}
		V_{0}^{n}=\lambda^{n}V_{0}=\sum_{i\geq0}(a_{i},\dots,w_{i})^{t}\lambda^{n-i},\quad V_{0,+}^{n}=\sum_{0}^{n}(a_{i},\dots,w_{i})^{t}\lambda^{n-i},\quad V_{0,-}^{n}=V_{0}^{n}-V_{0,+}^{n},
	\end{align*}
	then the zero curvature equation $\frac{\partial U_{0}}{\partial u}u_{t}+\frac{\partial U_{0}}{\partial\lambda}\lambda_{t}-V_{0,+x}^{n}+[U_{0},V_{0,+}^{n}]$ leads to the following Lax integrable hierarchy
	\begin{align*}
		u_{t_{n}}=\left(
		\begin{array}{c}
			u_{1}\\
			u_{2}\\
			u_{3}\\
			u_{4}\\
			u_{5}\\
			u_{6}\\
			u_{7}\\
			u_{8}\\
			u_{9}\\
			u_{10}\\
			u_{11}\\
			u_{12}
		\end{array}\right)_{t_{n}}=
		\left(
		\begin{array}{c}
			2k_{n+1}\\
			-2l_{n+1}\\
			2m_{n+1}\\
			-2o_{n+1}\\
			2p_{n+1}\\
			-2q_{n+1}\\
			2r_{n+1}\\
			-2s_{n+1}\\
			2t_{n+1}\\
			-2u_{n+1}\\
			2v_{n+1}\\
			-2w_{n+1}\\
		\end{array}\right)
	\end{align*}
	\begin{align}
		=\left(\begin{array}{cccccccccccc}
			0&1&0&0&0&0&0&0&0&0&0&0\\
			-1&0&0&0&0&0&0&0&0&0&0&0\\
			0&0&0&2&0&0&0&0&0&0&0&0\\
			0&0&-2&0&0&0&0&0&0&0&0&0\\
			0&0&0&0&0&2&0&0&0&0&0&0\\
			0&0&0&0&-2&0&0&0&0&0&0&0\\
			0&0&0&0&0&0&0&1&0&0&0&0\\
			0&0&0&0&0&0&-1&0&0&0&0&0\\
			0&0&0&0&0&0&0&0&0&1&0&0\\
			0&0&0&0&0&0&0&0&-1&0&0&0\\
			0&0&0&0&0&0&0&0&0&0&0&2\\
			0&0&0&0&0&0&0&0&0&0&-2&0
		\end{array}\right)
		\left(
		\begin{array}{c}
			2l_{n+1}\\
			2k_{n+1}\\
			o_{n+1}\\
			m_{n+1}\\
			q_{n+1}\\
			p_{n+1}\\
			2s_{n+1}\\
			2r_{n+1}\\
			2u_{n+1}\\
			2t_{n+1}\\
			w_{n+1}\\
			v_{n+1}\\
		\end{array}\right)=
		J_{1}P_{1,n+1}.\label{eq:2.4}
	\end{align}
 From the recurrence relations (\ref{eq:2.3}), we have
	\begin{align*}
		P_{1,n+1}=(l_{i,j})_{12\times12}\left(
		\begin{array}{c}
			2l_{n}\\
			2k_{n}\\
			o_{n}\\
			m_{n}\\
			q_{n}\\
			p_{n}\\
			2s_{n}\\
			2r_{n}\\
			2u_{n}\\
			2t_{n}\\
			w_{n}\\
			v_{n}\\
		\end{array}\right)+\left(
		\begin{array}{c}
		2u_{2}\\
		2u_{1}\\
		u_{4}\\u_{3}\\u_{6}\\u_{5}\\2u_{8}\\2u_{7}\\2u_{10}\\2u_{9}\\u_{12}\\u_{11}
		\end{array}\right)z_{n}(t)x=L_{1}P_{1,n}+\bar{u}z_{n}(t)x,
	\end{align*}
	where $\bar{u}=(2u_{2},2u_{1},u_{4},u_{3},u_{6},u_{5},2u_{8},2u_{7},2u_{10},2u_{9},u_{12},u_{11})^{t},$ and $L_{1}$ is a recurrence operator whose elements are given as follows
	\begin{align*}
&l_{1,1}=-\frac{1}{2}\partial+u_{2}\partial^{-1}u_{1}+\frac{1}{2}(u_{6}\partial^{-1}u_{5}+u_{8}\partial^{-1}u_{7}+u_{10}\partial^{-1}u_{9}+u_{12}\partial^{-1}u_{11}),\\ &l_{1,2}=-u_{2}\partial^{-1}u_{2}-\frac{1}{2}(u_{6}\partial^{-1}u_{12}+u_{12}\partial^{-1}u_{6}),\ l_{1,3}=0,\ l_{1,4}=-(u_{8}\partial^{-1}u_{10}+u_{10}\partial^{-1}u_{8}),\\
		&l_{1,5}=u_{2}\partial^{-1}u_{5}+u_{10}\partial^{-1}u_{7}+u_{12}\partial^{-1}u_{1},\ l_{1,6}=-(u_{2}\partial^{-1}u_{6}+u_{6}\partial^{-1}u_{2}),\\
		&l_{1,7}=\frac{1}{2}(u_{2}\partial^{-1}u_{7}+u_{10}\partial^{-1}u_{3}+u_{12}\partial^{-1}u_{9}),\ l_{1,8}=-\frac{1}{2}(u_{2}\partial^{-1}u_{8}+u_{6}\partial^{-1}u_{10}+u_{10}\partial^{-1}u_{6}+u_{8}\partial^{-1}u_{2}),\\
		&l_{1,9}=\frac{1}{2}(u_{2}\partial^{-1}u_{9}+u_{6}\partial^{-1}u_{7}+u_{8}\partial^{-1}u_{3}),\ 1_{1,10}=-\frac{1}{2}(u_{2}\partial^{-1}u_{10}+u_{8}\partial^{-1}u_{12}+u_{10}\partial^{-1}u_{2}+u_{12}\partial^{-1}u_{8}),\\
		&l_{1,11}=u_{2}\partial^{-1}u_{11}+u_{6}\partial^{-1}u_{1}+u_{8}\partial^{-1}u_{9},\ l_{1,12}=-(u_{2}\partial^{-1}u_{12}+u_{12}\partial^{-1}u_{2}),\\
		&l_{2,1}=u_{1}\partial^{-1}u_{1}+\frac{1}{2}(u_{5}\partial^{-1}u_{11}+u_{11}\partial^{-1}u_{5}),\\
		&l_{2,2}=\frac{1}{2}\partial-u_{1}\partial^{-1}u_{2}-\frac{1}{2}(u_{5}\partial^{-1}u_{6}+u_{7}\partial^{-1}u_{8}+u_{9}\partial^{-1}u_{10}+u_{11}\partial^{-1}u_{12}), \ l_{2,3}=u_{7}\partial^{-1}u_{9}+u_{9}\partial^{-1}u_{7},\\
		&l_{2,4}=0,\ l_{2,5}=u_{1}\partial^{-1}u_{5}+u_{5}\partial^{-1}u_{1},\ l_{2,6}=-(u_{1}\partial^{-1}u_{6}+u_{9}\partial^{-1}u_{8}+u_{11}\partial^{-1}u_{2}),\\
		&l_{2,7}=\frac{1}{2}(u_{1}\partial^{-1}u_{7}+u_{5}\partial^{-1}u_{9}+u_{7}\partial^{-1}u_{1}+u_{9}\partial^{-1}u_{5}),\ l_{2,8}=-\frac{1}{2}(u_{1}\partial^{-1}u_{8}+u_{9}\partial^{-1}u_{4}+u_{11}\partial^{-1}u_{10}),\\
		&l_{2,9}=\frac{1}{2}(u_{1}\partial^{-1}u_{9}+u_{7}\partial^{-1}u_{11}+u_{9}\partial^{-1}u_{1}+u_{11}\partial^{-1}u_{7}),\ l_{2,10}=-\frac{1}{2}(u_{1}\partial^{-1}u_{10}+u_{5}\partial^{-1}u_{8}+u_{7}\partial^{-1}u_{4}),\\
		&l_{2,11}=u_{1}\partial^{-1}u_{11}+u_{11}\partial^{-1}u_{1},\ l_{2,12}=-(u_{1}\partial^{-1}u_{12}+u_{5}\partial^{-1}u_{2}+u_{7}\partial^{-1}u_{10}),\ l_{3,1}=0,\\ &l_{3,2}=-\frac{1}{2}(u_{8}\partial^{-1}u_{10}+u_{10}\partial^{-1}u_{8}),\ l_{3,3}=-\frac{1}{2}\partial+(u_{4}\partial^{-1}u_{3}+u_{8}\partial^{-1}u_{7}+u_{10}\partial^{-1}u_{9}),\ l_{3,4}=-u_{4}\partial^{-1}u_{4},\\
		&l_{3,5}=0,\ l_{3,6}=-u_{8}\partial^{-1}u_{8},\ l_{3,7}=\frac{1}{2}(u_{4}\partial^{-1}u_{7}+u_{8}\partial^{-1}u_{4}+u_{10}\partial^{-1}u_{1}),\ l_{3,8}=-\frac{1}{2}(u_{4}\partial^{-1}u_{8}+u_{8}\partial^{-1}u_{4}),\\ &l_{3,9}=\frac{1}{2}(u_{4}\partial^{-1}u_{9}+u_{8}\partial^{-1}u_{1}+u_{10}\partial^{-1}u_{11}),\ l_{3,10}=-\frac{1}{2}(u_{4}\partial^{-1}u_{10}+u_{10}\partial^{-1}u_{4}),\ l_{3,11}=0,\\
		&l_{3,12}=-u_{10}\partial^{-1}u_{10},\ l_{4,1}=\frac{1}{2}(u_{7}\partial^{-1}u_{9}+u_{9}\partial^{-1}u_{7}),\ l_{4,2}=0,\ l_{4,3}=u_{3}\partial^{-1}u_{3},\\
		&l_{4,4}=\frac{1}{2}\partial-(u_{3}\partial^{-1}u_{4}+u_{7}\partial^{-1}u_{8}+u_{9}\partial^{-1}u_{10}),\ l_{4,5}=u_{7}\partial^{-1}u_{7},\ l_{4,6}=0,\ l_{4,7}=\frac{1}{2}(u_{3}\partial^{-1}u_{7}+u_{7}\partial^{-1}u_{3}),\\ &l_{4,8}=-\frac{1}{2}(u_{3}\partial^{-1}u_{8}+u_{7}\partial^{-1}u_{6}+u_{9}\partial^{-1}u_{2}),\ l_{4,9}=\frac{1}{2}(u_{3}\partial^{-1}u_{9}+u_{9}\partial^{-1}u_{3}),\\ &l_{4,10}=-\frac{1}{2}(u_{3}\partial^{-1}u_{10}+u_{7}\partial^{-1}u_{2}+u_{9}\partial^{-1}u_{12}),\ l_{4,11}=2u_{9}\partial^{-1}u_{9},\ l_{4,12}=0,\\
		&l_{5,1}=\frac{1}{2}(u_{6}\partial^{-1}u_{1}+u_{8}\partial^{-1}u_{9}+u_{2}\partial^{-1}u_{11}),\ l_{5,2}=-\frac{1}{2}(u_{6}\partial^{-1}u_{2}+u_{2}\partial^{-1}u_{6}),\ l_{5,3}=0,\ l_{5,4}=-u_{8}\partial^{-1}u_{8},\\ &l_{5,5}=-\frac{1}{2}\partial+u_{2}\partial^{-1}u_{1}+u_{6}\partial^{-1}u_{5}+u_{8}\partial^{-1}u_{7},\ l_{5,6}=-u_{6}\partial^{-1}u_{6},\ l_{5,7}=\frac{1}{2}(u_{2}\partial^{-1}u_{9}+u_{6}\partial^{-1}u_{7}+u_{8}\partial^{-1}u_{3}),\\
		&l_{5,8}=-\frac{1}{2}(u_{6}\partial^{-1}u_{8}+u_{8}\partial^{-1}u_{6}),\ l_{5,9}=0,\ l_{5,10}=-\frac{1}{2}(u_{2}\partial^{-1}u_{8}+u_{8}\partial^{-1}u_{2}),\ l_{5,11}=0,\ l_{5,12}=-u_{2}\partial^{-1}u_{2},\\
		&l_{6,1}=\frac{1}{2}(u_{1}\partial^{-1}u_{5}+u_{5}\partial^{-1}u_{1}),\ l_{6,2}=-\frac{1}{2}(u_{1}\partial^{-1}u_{12}+u_{5}\partial^{-1}u_{2}+u_{7}\partial^{-1}u_{10}),\ l_{6,3}=u_{7}\partial^{-1}u_{7},\ l_{6,4}=0,\\ &l_{6,5}=u_{5}\partial^{-1}u_{5},\ l_{6,6}=\frac{1}{2}\partial-u_{1}\partial^{-1}u_{2}+u_{5}\partial^{-1}u_{6}+u_{7}\partial^{-1}u_{8},\ l_{6,7}=\frac{1}{2}(u_{5}\partial^{-1}u_{7}+u_{7}\partial^{-1}u_{5}),\\ &l_{6,8}=-\frac{1}{2}(u_{1}\partial^{-1}u_{10}+u_{5}\partial^{-1}u_{8}+u_{7}\partial^{-1}u_{4}),\ l_{6,9}=\frac{1}{2}(u_{1}\partial^{-1}u_{7}+u_{7}\partial^{-1}u_{1}),\ l_{6,10}=0,\ l_{6,11}=u_{1}\partial^{-1}u_{1},\\
		&l_{6,12}=0,\ l_{7,1}=\frac{1}{2}(u_{4}\partial^{-1}u_{9}+u_{8}\partial^{-1}u_{1}+u_{10}\partial^{-1}u_{11}),\\ &l_{7,2}=-\frac{1}{2}(u_{2}\partial^{-1}u_{8}+u_{6}\partial^{-1}u_{10}+u_{8}\partial^{-1}u_{2}+u_{10}\partial^{-1}u_{6}),\ l_{7,3}=u_{2}\partial^{-1}u_{9}+u_{6}\partial^{-1}u_{7}+u_{8}\partial^{-1}u_{3},\\ &l_{7,4}=-(u_{4}\partial^{-1}u_{8}+u_{8}\partial^{-1}u_{4}),\ l_{7,5}=u_{4}\partial^{-1}u_{7}+u_{8}\partial^{-1}u_{5}+u_{10}\partial^{-1}u_{1},\ l_{7,6}=-(u_{6}\partial^{-1}u_{8}+u_{8}\partial^{-1}u_{6}),\\
		&l_{7,7}=-\frac{1}{2}\partial+u_{8}\partial^{-1}u_{7}+\frac{1}{2}(u_{2}\partial^{-1}u_{1}+u_{4}\partial^{-1}u_{3}+u_{6}\partial^{-1}u_{5}+u_{10}\partial^{-1}u_{9}),\\ &l_{7,8}=-u_{8}\partial^{-1}u_{8}-\frac{1}{2}(u_{4}\partial^{-1}u_{6}+u_{6}\partial^{-1}u_{4}),\ l_{7,9}=\frac{1}{2}(u_{2}\partial^{-1}u_{11}+u_{6}\partial^{-1}u_{1}+u_{8}\partial^{-1}u_{9}),\\ &l_{7,10}=-\frac{1}{2}(u_{2}\partial^{-1}u_{4}+u_{4}\partial^{-1}u_{2}+u_{8}\partial^{-1}u_{10}+u_{10}\partial^{-1}u_{8}),\ l_{7,11}=0,\ l_{7,12}=-(u_{2}\partial^{-1}u_{10}+u_{10}\partial^{-1}u_{2}),\\
		&l_{8,1}=\frac{1}{2}(u_{1}\partial^{-1}u_{7}+u_{5}\partial^{-1}u_{9}+u_{7}\partial^{-1}u_{1}+u_{9}\partial^{-1}u_{5}),\ l_{8,2}=-\frac{1}{2}(u_{3}\partial^{-1}u_{10}+u_{7}\partial^{-1}u_{2}+u_{9}\partial^{-1}u_{12}),\\
		&l_{8,3}=u_{3}\partial^{-1}u_{7}+u_{7}\partial^{-1}u_{3},\ l_{8,4}=-(u_{1}\partial^{-1}u_{10}+u_{5}\partial^{-1}u_{8}+u_{7}\partial^{-1}u_{4}),\ l_{8,5}=u_{5}\partial^{-1}u_{7}+u_{7}\partial^{-1}u_{5},\\ &l_{8,6}=-(u_{3}\partial^{-1}u_{8}+u_{7}\partial^{-1}u_{6}+u_{9}\partial^{-1}u_{2}),\ l_{8,7}=u_{7}\partial^{-1}u_{7}+\frac{1}{2}(u_{3}\partial^{-1}u_{5}+u_{5}\partial^{-1}u_{3}),\\
		&l_{8,8}=\frac{1}{2}\partial-u_{7}\partial^{-1}u_{8}-\frac{1}{2}(u_{1}\partial^{-1}u_{2}+u_{3}\partial^{-1}u_{4}+u_{5}\partial^{-1}u_{6}+u_{9}\partial^{-1}u_{10}),\\
		&l_{8,9}=\frac{1}{2}(u_{1}\partial^{-1}u_{3}+u_{3}\partial^{-1}u_{1}+u_{7}\partial^{-1}u_{9}+u_{9}\partial^{-1}u_{7}),\ l_{8,10}=-\frac{1}{2}(u_{1}\partial^{-1}u_{12}+u_{5}\partial^{-1}u_{2}+u_{7}\partial^{-1}u_{10}),\\
		&l_{8,11}=u_{1}\partial^{-1}u_{9}+u_{9}\partial^{-1}u_{1},\ l_{8,12}=0,\ l_{9,1}=\frac{1}{2}(u_{4}\partial^{-1}u_{7}+u_{8}\partial^{-1}u_{5}+u_{10}\partial^{-1}u_{1}),\\ &l_{9,2}=-\frac{1}{2}(u_{2}\partial^{-1}u_{10}+u_{8}\partial^{-1}u_{12}+u_{10}\partial^{-1}u_{2}+u_{12}\partial^{-1}u_{8}),\ l_{9,3}=u_{2}\partial^{-1}u_{7}+u_{10}\partial^{-1}u_{3}+u_{12}\partial^{-1}u_{9},\\ &l_{9,4}=-(u_{4}\partial^{-1}u_{10}+u_{10}\partial^{-1}u_{4}),\ l_{9,5}=0,\ l_{9,6}=-(u_{2}\partial^{-1}u_{8}+u_{8}\partial^{-1}u_{2}),\\ &l_{9,7}=\frac{1}{2}(u_{2}\partial^{-1}u_{5}+u_{10}\partial^{-1}u_{7}+u_{12}\partial^{-1}u_{1}),\ l_{9,8}=-\frac{1}{2}(u_{2}\partial^{-1}u_{4}+u_{4}\partial^{-1}u_{2}+u_{8}\partial^{-1}u_{10}+u_{10}\partial^{-1}u_{8}),\\
		&l_{9,9}=-\frac{1}{2}\partial+u_{10}\partial^{-1}u_{9}+\frac{1}{2}(u_{2}\partial^{-1}u_{1}+u_{4}\partial^{-1}u_{3}+u_{8}\partial^{-1}u_{7}+u_{12}\partial^{-1}u_{11}),\\
		&l_{9,10}=-u_{10}\partial^{-1}u_{10}-\frac{1}{2}(u_{4}\partial^{-1}u_{12}+u_{12}\partial^{-1}u_{4}),\ l_{9,11}=u_{4}\partial^{-1}u_{9}+u_{8}\partial^{-1}u_{1}+u_{10}\partial^{-1}u_{11},\\
		&l_{9,12}=-(u_{10}\partial^{-1}u_{12}+u_{12}\partial^{-1}u_{10}),\ l_{10,1}=\frac{1}{2}(u_{1}\partial^{-1}u_{9}+u_{7}\partial^{-1}u_{11}+u_{9}\partial^{-1}u_{1}+u_{11}\partial^{-1}u_{7}),\\
		&l_{10,2}=-\frac{1}{2}(u_{3}\partial^{-1}u_{8}+u_{7}\partial^{-1}u_{6}+u_{9}\partial^{-1}u_{2}),\ l_{10,3}=u_{3}\partial^{-1}u_{9}+u_{9}\partial^{-1}u_{3},\\
		&l_{10,4}=-(u_{1}\partial^{-1}u_{8}+u_{9}\partial^{-1}u_{4}+u_{11}\partial^{-1}u_{10}),\ l_{10,5}=u_{1}\partial^{-1}u_{7}+u_{7}\partial^{-1}u_{1},\ l_{10,6}=0,\\
		&l_{10,7}=\frac{1}{2}(u_{1}\partial^{-1}u_{3}+u_{3}\partial^{-1}u_{1}+u_{7}\partial^{-1}u_{9}+u_{9}\partial^{-1}u_{7}),\ l_{10,8}=-\frac{1}{2}(u_{1}\partial^{-1}u_{6}+u_{9}\partial^{-1}u_{8}+u_{11}\partial^{-1}u_{2}),\\
		&l_{10,9}=u_{9}\partial^{-1}u_{9}+\frac{1}{2}(u_{3}\partial^{-1}u_{11}+u_{11}\partial^{-1}u_{3}),\\
		&l_{10,10}=\frac{1}{2}\partial-u_{9}\partial^{-1}u_{10}-\frac{1}{2}(u_{1}\partial^{-1}u_{2}+u_{3}\partial^{-1}u_{4}+u_{7}\partial^{-1}u_{8}+u_{11}\partial^{-1}u_{12}),\\
		&l_{10,11}=u_{9}\partial^{-1}u_{11}+u_{11}\partial^{-1}u_{9},\ l_{10,12}=-(u_{3}\partial^{-1}u_{10}+u_{7}\partial^{-1}u_{2}+u_{9}\partial^{-1}u_{12}),\\
		&l_{11,1}=\frac{1}{2}(u_{2}\partial^{-1}u_{5}+u_{10}\partial^{-1}u_{7}+u_{12}\partial^{-1}u_{1}),\ l_{11,2}=-\frac{1}{2}(u_{2}\partial^{-1}u_{12}+u_{12}\partial^{-1}u_{2}),\ l_{11,3}=0,\\
		&l_{11,4}=-u_{10}\partial^{-1}u_{10},\ l_{11,5}=0,\ l_{11,6}=-u_{2}\partial^{-1}u_{2},\ l_{11,7}=0,\ l_{11,8}=-\frac{1}{2}(u_{2}\partial^{-1}u_{10}+u_{10}\partial^{-1}u_{2}),\\
		&l_{11,9}=\frac{1}{2}(u_{2}\partial^{-1}u_{7}+u_{10}\partial^{-1}u_{3}+u_{12}\partial^{-1}u_{9}),\ l_{11,10}=-\frac{1}{2}(u_{10}\partial^{-1}u_{12}+u_{12}\partial^{-1}u_{10}),\\
		&l_{11,11}=-\frac{1}{2}\partial+u_{2}\partial^{-1}u_{1}+u_{10}\partial^{-1}u_{9}+u_{12}\partial^{-1}u_{11},\ l_{11,12}=-u_{12}\partial^{-1}u_{12},\\ &l_{12,1}=\frac{1}{2}(u_{1}\partial^{-1}u_{11}+u_{11}\partial^{-1}u_{1}),\ l_{12,2}=-\frac{1}{2}(u_{1}\partial^{-1}u_{6}+u_{9}\partial^{-1}u_{8}+u_{11}\partial^{-1}u_{2}),\ l_{12,3}=u_{9}\partial^{-1}u_{9},\\
		&l_{12,4}=0,\ l_{12,5}=u_{1}\partial^{-1}u_{1},\ l_{12,6}=0,\ l_{12,7}=\frac{1}{2}(u_{1}\partial^{-1}u_{9}+u_{9}\partial^{-1}u_{1}),\ l_{12,8}=0,\\
		&l_{12,9}=\frac{1}{2}(u_{9}\partial^{-1}u_{11}+u_{11}\partial^{-1}u_{9}),\ l_{12,10}=-\frac{1}{2}(u_{1}\partial^{-1}u_{8}+u_{9}\partial^{-1}u_{4}+u_{11}\partial^{-1}u_{10}),\ l_{12,11}=u_{11}\partial^{-1}u_{11},\\
		&l_{12,12}=\frac{1}{2}\partial-(u_{1}\partial^{-1}u_{2}+u_{9}\partial^{-1}u_{10}+u_{11}\partial^{-1}u_{12}).
\end{align*}

We derive the Hamiltonian structure of (\ref{eq:2.4}) via the trace identity \cite{ref6},
\begin{align*}
	&\left\langle V_{0},\frac{\partial U_{0}}{\partial\lambda}\right\rangle = 2a+2b+2c ,\
	\left\langle V_{0},\frac{\partial U_{0}}{\partial u_{1}}\right\rangle = 2l ,\
	\left\langle V_{0},\frac{\partial U_{0}}{\partial u_{2}}\right\rangle = 2k ,\
	\left\langle V_{0},\frac{\partial U_{0}}{\partial u_{3}}\right\rangle = o ,\\
	&\left\langle V_{0},\frac{\partial U_{0}}{\partial u_{4}}\right\rangle = m ,\
	\left\langle V_{0},\frac{\partial U_{0}}{\partial u_{5}}\right\rangle = q ,\
	\left\langle V_{0},\frac{\partial U_{0}}{\partial u_{6}}\right\rangle = p ,\
	\left\langle V_{0},\frac{\partial U_{0}}{\partial u_{7}}\right\rangle = 2s ,\
	\left\langle V_{0},\frac{\partial U_{0}}{\partial u_{8}}\right\rangle = 2r ,\\
	&\left\langle V_{0},\frac{\partial U_{0}}{\partial u_{9}}\right\rangle = 2u ,\
	\left\langle V_{0},\frac{\partial U_{0}}{\partial u_{10}}\right\rangle = 2t ,\
	\left\langle V_{0},\frac{\partial U_{0}}{\partial u_{11}}\right\rangle = w ,\
	\left\langle V_{0},\frac{\partial U_{0}}{\partial u_{12}}\right\rangle = v .\
\end{align*}		

Substituting the above formulate into the trace identity yields
\begin{align*}
	\frac{\delta}{\delta u}(2a+2b+2c)=\lambda^{-\tau}\frac{\partial}{\partial\lambda}\lambda^{\tau}
	\left(\begin{array}{c}
		2l\\
		2k\\
		o\\
		m\\
		q\\
		p\\
		2s\\
		2r\\
		2u\\
		2t\\
		w\\
		v
	\end{array}\right),
\end{align*}
where $\tau=\frac{\lambda}{2}\frac{d}{dx}ln|tr(V_{0}^{2})|$. Balancing coefficients of each power of  in the above equality gives rise to
\begin{align*}
	\frac{\delta}{\delta u}(2a_{n+1}+2b_{n+1}+2c_{n+1})=(\tau-n)
	\left(\begin{array}{c}
		2l_{n}\\
		2k_{n}\\
		o_{n}\\
		m_{n}\\
		q_{n}\\
		p_{n}\\
		2s_{n}\\
		2r_{n}\\
		2u_{n}\\
		2t_{n}\\
		w_{n}\\
		v_{n}
	\end{array}\right).
\end{align*}
Taking $n=1$, gives $\tau=0$. Therefore we establish the following equation:
\begin{align*}
	P_{1,n+1}=
	\left(\begin{array}{c}
		2l_{n+1}\\
		2k_{n+1}\\
		o_{n+1}\\
		m_{n+1}\\
		q_{n+1}\\
		p_{n+1}\\
		2s_{n+1}\\
		2r_{n+1}\\
		2u_{n+1}\\
		2t_{n+1}\\
		w_{n+1}\\
		v_{n+1}
	\end{array}\right)=\frac{\delta}{\delta u}((\frac{-2}{n+1})(a_{n+2}+b_{n+2}+c_{n+2})).
\end{align*}
Thus, we see:
\begin{align*}
	u_{t}=J_{1}P_{1,n+1}=J_{1}\frac{\delta H_{n+1}^{1}}{\delta u}=J_{1}L_{1}\frac{\delta H_{n}^{1}}{\delta u}+J_{1}\bar{u}z_{n}(t)x,\\  H_{n+1}^{1}=(\frac{-2}{n+1})(a_{n+2}+b_{n+2}+c_{n+2}),\ n\geq0.
\end{align*}
When $n=1$, the hierarchy (\ref{eq:2.4}) reduces to the first integrable system
\begin{align*}
	u_{1t}=&\frac{1}{2}u_{1x}(\alpha(t)+\gamma(t))+\frac{1}{2}u_{9}\partial^{-1}(u_{1}u_{10}+u_{4}u_{7}+u_{5}u_{8})(\beta(t)-\alpha(t))\\
&+\frac{1}{2}u_{11}\partial^{-1}(u_{1}u_{12}+u_{2}u_{5}+u_{7}u_{10})(\gamma(t)-\alpha(t))\\
&+\frac{1}{2}u_{5}\partial^{-1}(u_{1}u_{6}+u_{2}u_{11}+u_{8}u_{9})(\alpha(t)-\gamma(t))\\
	&+\frac{1}{2}u_{7}\partial^{-1}(u_{1}u_{8}+u_{4}u_{9}+u_{10}u_{11})(\beta(t)-\gamma(t))+2u_{1}z_{1}(t)x,\\
	u_{2t}=&\frac{1}{2}u_{2x}(\alpha(t)+\gamma(t))-\frac{1}{2}u_{10}\partial^{-1}(u_{2}u_{9}+u_{3}u_{8}+u_{6}u_{7})(\alpha(t)-\beta(t))\\
	&-\frac{1}{2}u_{6}\partial^{-1}(u_{1}u_{12}+u_{2}u_{5}+u_{7}u_{10})(\gamma(t)-\alpha(t))-\frac{1}{2}u_{12}\partial^{-1}(u_{1}u_{6}+u_{2}u_{11}+u_{8}u_{9})(\alpha(t)-\gamma(t))\\
	&-\frac{1}{2}u_{8}\partial^{-1}(u_{2}u_{7}+u_{3}u_{10}+u_{9}u_{12})(\gamma(t)-\beta(t))-2u_{2}z_{1}(t)x,\\
	u_{3t}=&u_{3x}\beta(t)+u_{7}\partial^{-1}(u_{2}u_{9}+u_{3}u_{8}+u_{6}u_{7})(\alpha(t)-\beta(t))\\
	&+u_{9}\partial^{-1}(u_{2}u_{7}+u_{3}u_{10}+u_{9}u_{12})(\gamma(t)-\beta(t))+2u_{3}z_{1}(t)x,\\
	u_{4t}=&u_{4x}\beta(t)-u_{8}\partial^{-1}(u_{1}u_{10}+u_{4}u_{7}+u_{5}u_{8})(\beta(t)-\alpha(t))\\
	&-u_{10}\partial^{-1}(u_{1}u_{8}+u_{4}u_{9}+u_{10}u_{11})(\beta(t)-\gamma(t))-2u_{4}z_{1}(t)x,\\
	u_{5t}=&u_{5x}\alpha(t)+u_{7}\partial^{-1}(u_{1}u_{10}+u_{4}u_{7}+u_{5}u_{8})(\beta(t)-\alpha(t))\\
	&+u_{1}\partial^{-1}(u_{1}u_{12}+u_{2}u_{5}+u_{7}u_{10})(\gamma(t)-\alpha(t))+2u_{5}z_{1}(t)x,\\
	u_{6t}=&u_{6x}\alpha(t)-u_{8}\partial^{-1}(u_{2}u_{9}+u_{3}u_{8}+u_{6}u_{7})(\alpha(t)-\beta(t))\\
	&-u_{2}\partial^{-1}(u_{1}u_{6}+u_{2}u_{11}+u_{8}u_{9})(\alpha(t)-\gamma(t))-2u_{6}z_{1}(t)x,\\
	u_{7t}=&\frac{1}{2}u_{7x}(\alpha(t)+\beta(t))+\frac{1}{2}u_{3}\partial^{-1}(u_{1}u_{10}+u_{4}u_{7}+u_{5}u_{8})(\beta(t)-\alpha(t))\\
	&+\frac{1}{2}u_{5}\partial^{-1}(u_{2}u_{9}+u_{3}u_{8}+u_{6}u_{7})(\alpha(t)-\beta(t))+\frac{1}{2}u_{9}\partial^{-1}(u_{1}u_{12}+u_{2}u_{5}+u_{7}u_{10})(\gamma(t)-\alpha(t))\\
	&+\frac{1}{2}u_{1}\partial^{-1}(u_{2}u_{7}+u_{3}u_{10}+u_{9}u_{12})(\gamma(t)-\beta(t))+2u_{7}z_{1}(t)x,\\
	u_{8t}=&\frac{1}{2}u_{8x}(\alpha(t)+\beta(t))-\frac{1}{2}u_{6}\partial^{-1}(u_{1}u_{10}+u_{4}u_{7}+u_{5}u_{8})(\beta(t)-\alpha(t))\\
	&-\frac{1}{2}u_{4}\partial^{-1}(u_{2}u_{9}+u_{3}u_{8}+u_{6}u_{7})(\alpha(t)-\beta(t))-\frac{1}{2}u_{10}\partial^{-1}(u_{1}u_{6}+u_{2}u_{11}+u_{8}u_{9})(\alpha(t)-\gamma(t))\\
	&-\frac{1}{2}u_{2}\partial^{-1}(u_{1}u_{8}+u_{4}u_{9}+u_{10}u_{11})(\beta(t)-\gamma(t))-2u_{8}z_{1}(t)x,\\
	u_{9t}=&\frac{1}{2}u_{9x}(\beta(t)+\gamma(t))+\frac{1}{2}u_{1}\partial^{-1}(u_{2}u_{9}+u_{3}u_{8}+u_{6}u_{7})(\alpha(t)-\beta(t))\\
	&+\frac{1}{2}u_{7}\partial^{-1}(u_{1}u_{6}+u_{2}u_{11}+u_{8}u_{9})(\alpha(t)-\gamma(t))+\frac{1}{2}u_{11}\partial^{-1}(u_{2}u_{7}+u_{3}u_{10}+u_{9}u_{12})(\gamma(t)-\beta(t))\\
	&+\frac{1}{2}u_{3}\partial^{-1}(u_{1}u_{8}+u_{4}u_{9}+u_{10}u_{11})(\beta(t)-\gamma(t))+2u_{9}z_{1}(t)x,\\
	u_{10t}=&\frac{1}{2}u_{10x}(\beta(t)+\gamma(t))-\frac{1}{2}u_{2}\partial^{-1}(u_{1}u_{10}+u_{4}u_{7}+u_{5}u_{8})(\beta(t)-\alpha(t))\\
	&-\frac{1}{2}u_[8]\partial^{-1}(u_{1}u_{12}+u_{2}u_{5}+u_{7}u_{10})(\gamma(t)-\alpha(t))-\frac{1}{2}u_{4}\partial^{-1}(u_{2}u_{7}+u_{3}u_{10}+u_{9}u_{12})(\gamma(t)-\beta(t))\\
	&-\frac{1}{2}u_{12}\partial^{-1}(u_{1}u_{8}+u_{4}u_{9}+u_{10}u_{11})(\beta(t)-\gamma(t))-2u_{10}z_{1}(t)x,\\
	u_{11t}=&u_{11x}\gamma(t)+u_{1}\partial^{-1}(u_{1}u_{6}+u_{2}u_{11}+u_{8}u_{9})(\alpha(t)-\gamma(t))\\
	&+u_{9}\partial^{-1}(u_{1}u_{8}+u_{4}u_{9}+u_{10}u_{11})(\beta(t)-\gamma(t))+2u_{11}z_{1}(t)x,\\
	u_{12t}=&u_{12x}\gamma(t)-u_{2}\partial^{-1}(u_{1}u_{12}+u_{2}u_{5}+u_{7}u_{10})(\gamma(t)-\alpha(t))\\
	&-u_{10}\partial^{-1}(u_{2}u_{7}+u_{3}u_{10}+u_{9}u_{12})(\gamma(t)-\beta(t))-2u_{12}z_{1}(t)x.
\end{align*}
We construct a $(1+1)$-dimensional nonisospectral integrable hierarchy on the loop algebra of the symplectic Lie algebra $\mathfrak{sp}(6)$, and present the case of $n=1$ as an illustrative example.

\subsection{Soliton Hierarchy Associated with $\mathfrak{sp}(6)$ in (2+1) dimension.}\label{Sec:2-2}
Take
\begin{align*}
	\frac{\partial}{\partial z}=\frac{\partial}{\partial y}-\frac{\partial}{\partial x},\ \frac{\partial}{\partial z}=\frac{\partial}{\partial t}-\frac{\partial}{\partial x}
\end{align*}
and $U_{1},\;V_{1}\in\widetilde{\mathfrak{sp}(6)}$
\begin{align}
	U_{1}=\left(
	\begin{array}{cccccc}
		\lambda&0&0&u_{5}^{*}&u_{7}^{*}&u_{1}^{*}\\
		0&\lambda&0&u_{7}^{*}&u_{3}^{*}&u_{9}^{*}\\
		0&0&\lambda&u_{1}^{*}&u_{9}^{*}&u_{11}^{*}\\
		u_{6}^{*}&u_{8}^{*}&u_{2}^{*}&-\lambda&0&0\\
		u_{8}^{*}&u_{4}^{*}&u_{10}^{*}&0&-\lambda&0\\
		u_{2}^{*}&u_{10}^{*}&u_{12}^{*}&0&0&-\lambda
	\end{array}\right)\label{eq:3.1}
\end{align}
\begin{align}
	V_{1}=\left(
	\begin{array}{cccccc}
		a^{*}&d^{*}&f^{*}&p^{*}&r^{*}&k^{*}\\
		e^{*}&b^{*}&h^{*}&r^{*}&m^{*}&t^{*}\\
		g^{*}&j^{*}&c^{*}&k^{*}&t^{*}&v^{*}\\
		q^{*}&s^{*}&l^{*}&-a^{*}&-e^{*}&-g^{*}\\
		s^{*}&o^{*}&u^{*}&-d^{*}&-b^{*}&-j^{*}\\
		l^{*}&u^{*}&w^{*}&-f^{*}&-h^{*}&-c^{*}\\
	\end{array}\right)= \sum_{i\geq0}\left(
	\begin{array}{cccccc}
		a_{i}^{*}&d_{i}^{*}&f_{i}^{*}&p_{i}^{*}&r_{i}^{*}&k_{i}^{*}\\
		e_{i}^{*}&b_{i}^{*}&h_{i}^{*}&r_{i}^{*}&m_{i}^{*}&t_{i}^{*}\\
		g_{i}^{*}&j_{i}^{*}&c_{i}^{*}&k_{i}^{*}&t_{i}^{*}&v_{i}^{*}\\
		q_{i}^{*}&s_{i}^{*}&l_{i}^{*}&-a_{i}^{*}&-e_{i}^{*}&-g_{i}^{*}\\
		s_{i}^{*}&o_{i}^{*}&u_{i}^{*}&-d_{i}^{*}&-b_{i}^{*}&-j_{i}^{*}\\
		l_{i}^{*}&u_{i}^{*}&w_{i}^{*}&-f_{i}^{*}&-h_{i}^{*}&-c_{i}^{*}\\
	\end{array}\right)\lambda^{-i}.\label{eq:3.2}
\end{align}
Consider a nonisospectral problem in (2+1)-dimensions
\begin{align*}
	\begin{cases}
		\phi_{z}=U_{1}\phi\\
		\phi_{w}=V_{1}\phi\\
		\lambda_{t}=\sum_{i\geq0}z_{i}^{*}(t)\lambda^{-i}.
	\end{cases}
\end{align*}
The zero curvature equation $U_{1w}-V_{1z}+[U_{1},V_{1}]=0$ can be rewritten in (2+1)-dimensional form $U_{1t}-U_{1x}+V_{1x}-V_{1y}+[U_{1},V_{1}]=0$, the stationary zero curvature representation $V_{1z}=\frac{\partial U}{\partial\lambda}\partial_{t}+[U_{1},V_{1}]$ can be rewritten as $V_{1y}-V_{1x}=\frac{\partial U}{\partial\lambda}\partial_{t}+[U_{1},V_{1}]$, and satisfies the compatibility condition for the following Lax pair
\begin{align*}
	\begin{cases}
		\phi_{y}=\phi_{x}+U_{1}\phi\\
		\phi_{t}=\phi_{x}+V_{1}\phi
	\end{cases}
\end{align*}
The stationary zero curvature representation $V_{1y}-V_{1x}=\frac{\partial U}{\partial\lambda}\partial_{t}+[U_{1},V_{1}]$ gives
\begin{align}
	\begin{cases}
		a_{x}^{*}=-u_{1}^{*}l^{*}+u_{2}^{*}k^{*}-u_{5}^{*}q^{*}+u_{6}^{*}p^{*}-u_{7}^{*}s^{*}+u_{8}^{*}r^{*}-z(t)+a_{y}^{*}\\
		b_{x}^{*}=-u_{3}^{*}o^{*}+u_{4}^{*}m^{*}-u_{7}^{*}s^{*}+u_{8}^{*}r^{*}-u_{9}^{*}u^{*}+u_{10}^{*}t^{*}-z(t)+b_{y}^{*}\\
		c_{x}^{*}=-u_{1}^{*}l^{*}+u_{2}^{*}k^{*}-u_{9}^{*}u^{*}+u_{10}^{*}t^{*}-u_{11}^{*}w^{*}+u_{12}^{*}v^{*}-z(t)+c_{y}^{*}\\
		d_{x}^{*}=-u_{1}^{*}u^{*}+u_{4}^{*}r^{*}-u_{5}^{*}s^{*}-u_{7}^{*}o^{*}+u_{8}^{*}p^{*}+u_{10}^{*}k^{*}+d_{y}^{*}\\
		e_{x}^{*}=u_{2}^{*}t^{*}-u_{3}^{*}s^{*}+u_{6}^{*}r^{*}-u_{7}^{*}q^{*}+u_{8}^{*}m^{*}-u_{9}^{*}l^{*}+e_{y}^{*}\\
		f_{x}^{*}=-u_{1}^{*}w^{*}+u_{2}^{*}p^{*}-u_{5}^{*}l^{*}-u_{7}^{*}u^{*}+u_{10}^{*}r^{*}+u_{12}^{*}k^{*}+f_{y}^{*}\\
		g_{x}^{*}=-u_{1}^{*}q^{*}+u_{2}^{*}v^{*}+u_{6}^{*}k^{*}+u_{8}^{*}t^{*}-u_{9}^{*}s^{*}-u_{11}^{*}l^{*}+g_{y}^{*}\\
		h_{x}^{*}=u_{2}^{*}r^{*}-u_{3}^{*}u^{*}-u_{7}^{*}l^{*}-u_{9}^{*}w^{*}+u_{10}^{*}m^{*}+u_{12}^{*}t^{*}+h_{y}^{*}\\
		j_{x}^{*}=-u_{1}^{*}s^{*}+u_{4}^{*}t^{*}+u_{8}^{*}k^{*}-u_{9}^{*}o^{*}+u_{10}^{*}v^{*}-u_{11}^{*}u^{*}+j_{y}^{*}\\
		k_{x}^{*}=-2\lambda k^{*}+u_{1}^{*}a^{*}+u_{1}^{*}c^{*}+u_{5}^{*}g^{*}+u_{7}^{*}j^{*}+u_{9}^{*}d^{*}+u_{11}^{*}f^{*}+k_{y}^{*}\\
		l_{x}^{*}=2\lambda l^{*}-u_{2}^{*}a^{*}-u_{2}^{*}c^{*}-u_{6}^{*}f^{*}-u_{8}^{*}h^{*}-u_{10}^{*}e^{*}-u_{12}^{*}g^{*}+l_{y}^{*}\\
		m_{x}^{*}=-2\lambda m^{*}+2u_{3}^{*}b^{*}+2u_{7}^{*}e^{*}+2u_{9}^{*}h^{*}+m_{y}^{*}\\
		o_{x}^{*}=2\lambda o^{*}-2u_{4}^{*}b^{*}-2u_{8}^{*}d^{*}-2u_{10}^{*}j^{*}+o_{y}^{*}\\
		p_{x}^{*}=-2\lambda p^{*}+2u_{1}^{*}f^{*}+2u_{5}^{*}a^{*}+2u_{7}^{*}d^{*}+p_{y}^{*}\\
		q_{x}^{*}=2\lambda q^{*}-2u_{2}^{*}g^{*}-2u_{6}^{*}a^{*}-2u_{8}^{*}e^{*}+q_{y}^{*}\\
		r_{x}^{*}=-2\lambda r^{*}+u_{1}^{*}h^{*}+u_{3}^{*}d^{*}+u_{5}^{*}e^{*}+u_{7}^{*}a^{*}+u_{7}^{*}b^{*}+u_{9}^{*}f^{*}+r_{y}^{*}\\
		s_{x}^{*}=2\lambda s^{*}-u_{2}^{*}j^{*}-u_{4}^{*}e^{*}-u_{6}^{*}d^{*}-u_{8}^{*}a^{*}-u_{8}^{*}b^{*}-u_{10}^{*}g^{*}+s_{y}^{*}\\
		t_{x}^{*}=-2\lambda t^{*}+u_{1}^{*}e^{*}+u_{3}^{*}j^{*}+u_{7}^{*}g^{*}+u_{9}^{*}b^{*}+u_{9}^{*}c^{*}+u_{11}^{*}h^{*}+t_{y}^{*}\\
		u_{x}^{*}=2\lambda u^{*}-u_{2}^{*}d^{*}-u_{4}^{*}h^{*}-u_{8}^{*}f^{*}-u_{10}^{*}b^{*}-u_{10}^{*}c^{*}-u_{12}^{*}j^{*}+u_{y}^{*}\\
		v_{x}^{*}=-2\lambda v^{*}+2u_{1}^{*}g^{*}+2u_{9}^{*}j^{*}+2u_{11}^{*}c^{*}+v_{y}^{*}\\
		w_{x}^{*}=2\lambda w^{*}-2u_{2}^{*}f^{*}-2u_{10}^{*}h^{*}-2u_{12}^{*}c^{*}+w_{y}^{*}
	\end{cases}\label{eq:2.7}
\end{align}
	Take the initial values
\begin{align*}
	a_{0}^{*}=\alpha(t), b_{0}^{*}=\beta(t), c_{0}^{*}=\gamma(t), d_{0}^{*}=e_{0}^{*}=\dots=v_{0}^{*}=w_{0}^{*}=z_{0}(t)=0,
\end{align*}
then we have
\begin{align*}
	a_{1}^{*}=&-z_{1}(t)x+\partial_{x}^{-1}a_{1y}^{*},\ b_{1}^{*}=-z_{1}(t)x+\partial_{x}^{-1}b_{1y}^{*},\ c_{1}^{*}=-z_{1}(t)x+\partial_{x}^{-1}c_{1y}^{*},\\ d_{1}^{*}=&\frac{1}{2}\partial_{x}^{-1}(u_{1}^{*}u_{10}^{*}+u_{4}^{*}u_{7}^{*}+u_{5}^{*}u_{8}^{*})(\beta(t)-\alpha(t))+\partial_{x}^{-1}d_{1y}^{*},\\
	e_{1}^{*}=&\frac{1}{2}\partial_{x}^{-1}(u_{2}^{*}u_{9}^{*}+u_{3}^{*}u_{8}^{*}+u_{6}^{*}u_{7}^{*})(\alpha(t)-\beta(t))+\partial_{x}^{-1}e_{1y}^{*},\\ f_{1}^{*}=&\frac{1}{2}\partial_{x}^{-1}(u_{1}^{*}u_{12}^{*}+u_{2}^{*}u_{5}^{*}+u_{7}^{*}u_{10}^{*})(\gamma(t)-\alpha(t))+\partial_{x}^{-1}f_{1y}^{*},\\
	g_{1}^{*}=&\frac{1}{2}\partial_{x}^{-1}(u_{1}^{*}u_{6}^{*}+u_{2}^{*}u_{11}^{*}+u_{8}^{*}u_{9}^{*})(\alpha(t)-\gamma(t))+\partial_{x}^{-1}g_{1y}^{*},\\
	h_{1}^{*}=&\frac{1}{2}\partial_{x}^{-1}(u_{2}^{*}u_{7}^{*}+u_{3}^{*}u_{10}^{*}+u_{9}^{*}u_{12}^{*})(\gamma(t)-\beta(t))+\partial_{x}^{-1}h_{1y}^{*},\\
	j_{1}^{*}=&\frac{1}{2}\partial_{x}^{-1}(u_{1}^{*}u_{8}^{*}+u_{4}^{*}u_{9}^{*}+u_{10}^{*}u_{11}^{*})(\beta(t)-\gamma(t))+\partial_{x}^{-1}j_{1y}^{*},\\
	k_{1}^{*}=&\frac{1}{2}u_{1}^{*}(\alpha(t)+\gamma(t)),\ l_{1}^{*}=\frac{1}{2}u_{2}^{*}(\alpha(t)+\gamma(t)),\\
	m_{1}^{*}=&u_{3}^{*}\beta(t),\ o_{1}^{*}=u_{4}^{*}\beta(t),\ p_{1}^{*}=u_{5}^{*}\alpha(t),\ q_{1}^{*}=u_{6}^{*}\alpha(t),\ r_{1}^{*}=\frac{1}{2}u_{7}^{*}(\alpha(t)+\beta(t)),\ s_{1}^{*}=\frac{1}{2}u_{8}^{*}(\alpha(t)+\beta(t)),\\
	t_{1}^{*}=&\frac{1}{2}u_{9}^{*}(\beta(t)+\gamma(t)),\ u_{1}^{*}=\frac{1}{2}u_{10}^{*}(\beta(t)+\gamma(t)),\
	v_{1}^{*}=u_{11}^{*}\gamma(t),\ w_{1}^{*}=u_{12}^{*}\gamma(t),\\
	k_{2}^{*}=&\frac{1}{4}u_{1x}^{*}(\alpha(t)+\gamma(t))+\frac{1}{4}u_{5}^{*}\partial^{-1}(u_{1}^{*}u_{6}^{*}+u_{2}^{*}u_{11}^{*}+u_{8}^{*}u_{9}^{*})(\alpha(t)-\gamma(t))\\
	&+\frac{1}{4}u_{7}^{*}\partial^{-1}(u_{1}^{*}u_{8}^{*}+u_{4}^{*}u_{9}^{*}+u_{10}^{*}u_{11}^{*})(\beta(t)-\gamma(t))+\frac{1}{4}u_{9}^{*}\partial^{-1}(u_{1}^{*}u_{10}^{*}+u_{4}^{*}u_{7}^{*}+u_{5}^{*}u_{8}^{*})(\beta(t)-\alpha(t))\\
	&+\frac{1}{4}u_{11}^{*}\partial^{-1}(u_{1}^{*}u_{12}^{*}+u_{2}^{*}u_{5}^{*}+u_{7}^{*}u_{10}^{*})(\gamma(t)-\alpha(t))-u_{1}^{*}z_{1}(t)x\\
	&+\frac{1}{2}(k_{1y}^{*}+u_{1}^{*}\partial_{x}^{-1}a_{1y}^{*}+u_{1}^{*}\partial_{x}^{-1}c_{1y}^{*}+u_{5}^{*}\partial_{x}^{-1}g_{1y}^{*}+u_{7}^{*}\partial_{x}^{-1}j_{1y}^{*}+u_{9}^{*}\partial_{x}^{-1}d_{1y}^{*}+u_{11}^{*}\partial_{x}^{-1}f_{1y}^{*}),\\
	l_{2}^{*}=&-\frac{1}{4}u_{2x}^{*}(\alpha(t)+\gamma(t))+\frac{1}{4}u_{6}^{*}\partial^{-1}(u_{1}^{*}u_{12}^{*}+u_{2}^{*}u_{5}^{*}+u_{7}^{*}u_{10}^{*})(\gamma(t)-\alpha(t))\\
	&+\frac{1}{4}u_{8}^{*}\partial^{-1}(u_{2}^{*}u_{7}^{*}+u_{3}^{*}u_{10}^{*}+u_{9}^{*}u_{12}^{*})(\gamma(t)-\beta(t))+\frac{1}{4}u_{10}^{*}\partial^{-1}(u_{2}^{*}u_{9}^{*}+u_{3}^{*}u_{8}^{*}+u_{6}^{*}u_{7}^{*})(\alpha(t)-\beta(t))\\
	&+\frac{1}{4}u_{12}^{*}\partial^{-1}(u_{1}^{*}u_{6}^{*}+u_{2}^{*}u_{11}^{*}+u_{8}^{*}u_{9}^{*})(\alpha(t)-\gamma(t))-u_{2}^{*}z_{1}(t)x\\
	&+\frac{1}{2}(-l_{1y}^{*}+u_{2}^{*}\partial_{x}^{-1}a_{1y}^{*}+u_{2}^{*}\partial_{x}^{-1}c_{1y}^{*}+u_{6}^{*}\partial_{x}^{-1}f_{1y}^{*}+u_{8}^{*}\partial_{x}^{-1}h_{1y}^{*}+u_{10}^{*}\partial_{x}^{-1}e_{1y}^{*}+u_{12}^{*}\partial_{x}^{-1}g_{1y}^{*}),\\
	m_{2}^{*}=&\frac{1}{2}u_{3x}^{*}\beta(t)+\frac{1}{2}u_{7}^{*}\partial^{-1}(u_{2}^{*}u_{9}^{*}+u_{3}^{*}u_{8}^{*}+u_{6}^{*}u_{7}^{*})(\alpha(t)-\beta(t))\\
	&+\frac{1}{2}u_{9}^{*}\partial^{-1}(u_{2}^{*}u_{7}^{*}+u_{3}^{*}u_{10}^{*}+u_{9}^{*}u_{12}^{*})(\gamma(t)-\beta(t))-u_{3}^{*}z_{1}(t)x\\
	&+\frac{1}{2}(m_{1y}^{*}+2u_{3}^{*}\partial_{x}^{-1}b_{1y}^{*}+2u_{7}^{*}\partial_{x}^{-1}e_{1y}^{*}+2u_{9}^{*}\partial_{x}^{-1}h_{1y}^{*}),\\
	o_{2}^{*}=&-\frac{1}{2}u_{4x}^{*}\beta(t)+\frac{1}{2}u_{8}^{*}\partial^{-1}(u_{1}^{*}u_{10}^{*}+u_{4}^{*}u_{7}^{*}+u_{5}^{*}u_{8}^{*})(\beta(t)-\alpha(t))\\
	&+\frac{1}{2}u_{10}^{*}\partial^{-1}(u_{1}^{*}u_{8}^{*}+u_{4}^{*}u_{9}^{*}+u_{10}^{*}u_{11}^{*})(\beta(t)-\gamma(t))-u_{4}^{*}z_{1}(t)x\\
	&+\frac{1}{2}(-o_{1y}^{*}+2u_{4}^{*}\partial_{x}^{-1}b_{1y}^{*}+u_{8}^{*}\partial_{x}^{-1}d_{1y}^{*}+u_{10}^{*}\partial_{x}^{-1}j_{1y}^{*}),\\
	p_{2}^{*}=&\frac{1}{2}u_{5x}^{*}\alpha(t)+\frac{1}{2}u_{1}^{*}\partial^{-1}(u_{1}^{*}u_{12}^{*}+u_{2}^{*}u_{5}^{*}+u_{7}^{*}u_{10}^{*})(\gamma(t)-\alpha(t))\\
	&+\frac{1}{2}u_{7}^{*}\partial^{-1}(u_{1}^{*}u_{10}^{*}+u_{4}^{*}u_{7}^{*}+u_{5}^{*}u_{8}^{*})(\beta(t)-\alpha(t))-u_{5}^{*}z_{1}(t)x\\
	&+\frac{1}{2}(p_{1y}^{*}+2u_{1}^{*}\partial_{x}^{-1}f_{1y}^{*}+2u_{5}^{*}\partial_{x}^{-1}a_{1y}^{*}+2u_{7}^{*}\partial_{x}^{-1}d_{1y}^{*}),\\
	q_{2}^{*}=&-\frac{1}{2}u_{6x}^{*}\alpha(t)+\frac{1}{2}u_{2}^{*}\partial^{-1}(u_{1}^{*}u_{6}^{*}+u_{2}^{*}u_{11}^{*}+u_{8}^{*}u_{9}^{*})(\alpha(t)-\gamma(t))\\
	&+\frac{1}{2}u_{8}^{*}\partial^{-1}(u_{2}^{*}u_{9}^{*}+u_{3}^{*}u_{8}^{*}+u_{6}^{*}u_{7}^{*})(\alpha(t)-\beta(t))-u_{6}^{*}z_{1}(t)x\\
	&+\frac{1}{2}(-q_{1y}^{*}+2u_{2}^{*}\partial_{x}^{-1}g_{1y}^{*}+2u_{6}^{*}\partial_{x}^{-1}a_{1y}^{*}+2u_{8}^{*}\partial_{x}^{-1}e_{1y}^{*}),\\
	r_{2}^{*}=&\frac{1}{4}u_{7x}^{*}(\alpha(t)+\beta(t))+\frac{1}{4}u_{1}^{*}\partial^{-1}(u_{2}^{*}u_{7}^{*}+u_{3}^{*}u_{10}^{*}+u_{9}^{*}u_{12}^{*})(\gamma(t)-\beta(t))\\
	&+\frac{1}{4}u_{3}^{*}\partial^{-1}(u_{1}^{*}u_{10}^{*}+u_{4}^{*}u_{7}^{*}+u_{5}^{*}u_{8}^{*})(\beta(t)-\alpha(t))+\frac{1}{4}u_{5}^{*}\partial^{-1}(u_{2}^{*}u_{9}^{*}+u_{3}^{*}u_{8}^{*}+u_{6}^{*}u_{7}^{*})(\alpha(t)-\beta(t))\\
	&+\frac{1}{4}u_{9}^{*}\partial^{-1}(u_{1}^{*}u_{12}^{*}+u_{2}^{*}u_{5}^{*}+u_{7}^{*}u_{10}^{*})(\gamma(t)-\alpha(t))-u_{7}^{*}z_{1}(t)x\\
	&+\frac{1}{2}(r_{1y}^{*}+u_{1}^{*}\partial_{x}^{-1}h_{1y}^{*}+u_{3}^{*}\partial_{x}^{-1}d_{1y}^{*}+u_{5}^{*}\partial_{x}^{-1}e_{1y}^{*}+u_{7}^{*}\partial_{x}^{-1}a_{1y}^{*}+u_{7}^{*}\partial_{x}^{-1}b_{1y}^{*}+u_{9}^{*}\partial_{x}^{-1}f_{1y}^{*}),\\
	s_{2}^{*}=&-\frac{1}{4}u_{8x}^{*}(\alpha(t)+\beta(t))+\frac{1}{4}u_{2}^{*}\partial^{-1}(u_{1}^{*}u_{8}^{*}+u_{4}^{*}u_{9}^{*}+u_{10}^{*}u_{11}^{*})(\beta(t)-\gamma(t))\\
	&+\frac{1}{4}u_{4}^{*}\partial^{-1}(u_{2}^{*}u_{9}^{*}+u_{3}^{*}u_{8}^{*}+u_{6}^{*}u_{7}^{*})(\alpha(t)-\beta(t))+\frac{1}{4}u_{6}^{*}\partial^{-1}(u_{1}^{*}u_{10}^{*}+u_{4}^{*}u_{7}^{*}+u_{5}^{*}u_{8}^{*})(\beta(t)-\alpha(t))\\
	&+\frac{1}{4}u_{10}^{*}\partial^{-1}(u_{1}^{*}u_{6}^{*}+u_{2}^{*}u_{11}^{*}+u_{8}^{*}u_{9}^{*})(\alpha(t)-\gamma(t))-u_{8}^{*}z_{1}(t)x\\
	&+\frac{1}{2}(-s_{1y}^{*}+u_{2}^{*}\partial_{x}^{-1}j_{1y}^{*}+u_{4}^{*}\partial_{x}^{-1}e_{1y}^{*}+u_{6}^{*}\partial_{x}^{-1}d_{1y}^{*}+u_{8}^{*}\partial_{x}^{-1}a_{1y}^{*}+u_{8}^{*}\partial_{x}^{-1}b_{1y}^{*}+u_{10}^{*}\partial_{x}^{-1}g_{1y}^{*}),\\
	t_{2}^{*}=&\frac{1}{4}u_{9x}^{*}(\beta(t)+\gamma(t))+\frac{1}{4}u_{1}^{*}\partial^{-1}(u_{2}^{*}u_{9}^{*}+u_{3}^{*}u_{8}^{*}+u_{6}^{*}u_{7}^{*})(\alpha(t)-\beta(t))\\
	&+\frac{1}{4}u_{3}^{*}\partial^{-1}(u_{1}^{*}u_{8}^{*}+u_{4}^{*}u_{9}^{*}+u_{10}^{*}u_{11}^{*})(\beta(t)-\gamma(t))+\frac{1}{4}u_{7}^{*}\partial^{-1}(u_{1}^{*}u_{6}^{*}+u_{2}^{*}u_{11}^{*}+u_{8}^{*}u_{9}^{*})(\alpha(t)-\gamma(t))\\
	&+\frac{1}{4}u_{11}^{*}\partial^{-1}(u_{2}^{*}u_{7}^{*}+u_{3}^{*}u_{10}^{*}+u_{9}^{*}u_{12}^{*})(\gamma(t)-\beta(t))-u_{9}^{*}z_{1}(t)x\\
	&+\frac{1}{2}(t_{1y}^{*}+u_{1}^{*}\partial_{x}^{-1}e_{1y}^{*}+u_{3}^{*}\partial_{x}^{-1}j_{1y}^{*}+u_{7}^{*}\partial_{x}^{-1}g_{1y}^{*}+u_{9}^{*}\partial_{x}^{-1}b_{1y}^{*}+u_{9}^{*}\partial_{x}^{-1}c_{1y}^{*}+u_{11}^{*}\partial_{x}^{-1}h_{1y}^{*}),\\
	u_{2}^{*}=&-\frac{1}{4}u_{10x}^{*}(\beta(t)+\gamma(t))+\frac{1}{4}u_{2}^{*}\partial^{-1}(u_{1}^{*}u_{10}^{*}+u_{4}^{*}u_{7}^{*}+u_{5}^{*}u_{8}^{*})(\beta(t)-\alpha(t))\\
	&+\frac{1}{4}u_{4}^{*}\partial^{-1}(u_{2}^{*}u_{7}^{*}+u_{3}^{*}u_{10}^{*}+u_{9}^{*}u_{12}^{*})(\gamma(t)-\beta(t))+\frac{1}{4}u_{8}^{*}\partial^{-1}(u_{1}^{*}u_{12}^{*}+u_{2}^{*}u_{5}^{*}+u_{7}^{*}u_{10}^{*})(\gamma(t)-\alpha(t))\\
	&+\frac{1}{4}u_{12}^{*}\partial^{-1}(u_{1}^{*}u_{8}^{*}+u_{4}^{*}u_{9}^{*}+u_{10}^{*}u_{11}^{*})(\beta(t)-\gamma(t))-u_{10}^{*}z_{1}(t)x\\
	&+\frac{1}{2}(-u_{1y}^{*}+u_{2}^{*}\partial_{x}^{-1}d_{1y}^{*}+u_{4}^{*}\partial_{x}^{-1}h_{1y}^{*}+u_{8}^{*}\partial_{x}^{-1}f_{1y}^{*}+u_{10}^{*}\partial_{x}^{-1}b_{1y}^{*}+u_{10}^{*}\partial_{x}^{-1}c_{1y}^{*}+u_{12}^{*}\partial_{x}^{-1}j_{1y}^{*}),\\
	v_{2}^{*}=&\frac{1}{2}u_{11x}^{*}\gamma(t)+\frac{1}{2}u_{1}^{*}\partial^{-1}(u_{1}^{*}u_{6}^{*}+u_{2}^{*}u_{11}^{*}+u_{8}^{*}u_{9}^{*})(\alpha(t)-\gamma(t))\\
	&+\frac{1}{2}u_{9}^{*}\partial^{-1}(u_{1}^{*}u_{8}^{*}+u_{4}^{*}u_{9}^{*}+u_{10}^{*}u_{11}^{*})(\beta(t)-\gamma(t))-u_{11}^{*}z_{1}(t)x\\
	&+\frac{1}{2}(v_{1y}^{*}+2u_{1}^{*}\partial_{x}^{-1}g_{1y}^{*}+2u_{9}^{*}\partial_{x}^{-1}j_{1y}^{*}+2u_{11}^{*}\partial_{x}^{-1}c_{1y}^{*}),\\
	w_{2}^{*}=&-\frac{1}{2}u_{12x}^{*}\gamma(t)+\frac{1}{2}u_{2}^{*}\partial^{-1}(u_{1}^{*}u_{12}^{*}+u_{2}^{*}u_{5}^{*}+u_{7}^{*}u_{10}^{*})(\gamma(t)-\alpha(t))\\
	&+\frac{1}{2}u_{10}^{*}\partial^{-1}(u_{2}^{*}u_{7}^{*}+u_{3}^{*}u_{10}^{*}+u_{9}^{*}u_{12}^{*})(\gamma(t)-\beta(t))-u_{12}^{*}z_{1}(t)x\\
	&+\frac{1}{2}(-w_{1y}^{*}+2u_{2}^{*}\partial_{x}^{-1}f_{1y}^{*}+2u_{10}^{*}\partial_{x}^{-1}h_{1y}^{*}+2u_{12}^{*}\partial_{x}^{-1}c_{1y}^{*}),\\
	\dots&\dots
\end{align*}
Now, taking
\begin{align*}
	V_{1}^{n}=\lambda^{n}V_{1}=\sum_{i\geq0}(a_{i}^{*},\dots,w_{i}^{*})^{t}\lambda^{n-i},\quad V_{1,+}^{n}=\sum_{0}^{n}(a_{i}^{*},\dots,w_{i}^{*})^{t}\lambda^{n-i},\quad V_{1,-}^{n}=V_{1}^{n}-V_{1,+}^{n},
\end{align*}
then the zero curvature equation
$U_{1t}-U_{1x}+V_{1,+x}^{(n)}-V_{1,+y}^{(n)}+[U_{1},V_{1,+}^{(n)}]=0$ leads to the following Lax integrable hierarchy
\begin{align*}
	u_{t_{n}}^{*}=\left(
	\begin{array}{c}
		u_{1}^{*}\\
		u_{2}^{*}\\
		u_{3}^{*}\\
		u_{4}^{*}\\
		u_{5}^{*}\\
		u_{6}^{*}\\
		u_{7}^{*}\\
		u_{8}^{*}\\
		u_{9}^{*}\\
		u_{10}^{*}\\
		u_{11}^{*}\\
		u_{12}^{*}
	\end{array}\right)_{t_{n}}=
	\left(
	\begin{array}{c}
		2k_{n+1}^{*}+u_{1x}^{*}\\
		-2l_{n+1}^{*}+u_{2x}^{*}\\
		2m_{n+1}^{*}+u_{3x}^{*}\\
		-2o_{n+1}^{*}+u_{4x}^{*}\\
		2p_{n+1}^{*}+u_{5x}^{*}\\
		-2q_{n+1}^{*}+u_{6x}^{*}\\
		2r_{n+1}^{*}+u_{7x}^{*}\\
		-2s_{n+1}^{*}+u_{8x}^{*}\\
		2t_{n+1}^{*}+u_{9x}^{*}\\
		-2u_{n+1}^{*}+u_{10x}^{*}\\
		2v_{n+1}^{*}+u_{11x}^{*}\\
		-2w_{n+1}^{*}+u_{12x}^{*}\\
	\end{array}\right)
\end{align*}
\begin{align}
	=\left(\begin{array}{cccccccccccc}
		0&1&0&0&0&0&0&0&0&0&0&0\\
		-1&0&0&0&0&0&0&0&0&0&0&0\\
		0&0&0&2&0&0&0&0&0&0&0&0\\
		0&0&-2&0&0&0&0&0&0&0&0&0\\
		0&0&0&0&0&2&0&0&0&0&0&0\\
		0&0&0&0&-2&0&0&0&0&0&0&0\\
		0&0&0&0&0&0&0&1&0&0&0&0\\
		0&0&0&0&0&0&-1&0&0&0&0&0\\
		0&0&0&0&0&0&0&0&0&1&0&0\\
		0&0&0&0&0&0&0&0&-1&0&0&0\\
		0&0&0&0&0&0&0&0&0&0&0&2\\
		0&0&0&0&0&0&0&0&0&0&-2&0
	\end{array}\right)
	\left(
	\begin{array}{c}
		2l_{n+1}^{*}\\
		2k_{n+1}^{*}\\
		o_{n+1}^{*}\\
		m_{n+1}^{*}\\
		q_{n+1}^{*}\\
		p_{n+1}^{*}\\
		2s_{n+1}^{*}\\
		2r_{n+1}^{*}\\
		2u_{n+1}^{*}\\
		2t_{n+1}^{*}\\
		w_{n+1}^{*}\\
		v_{n+1}^{*}\\
	\end{array}\right)+
	\left(\begin{array}{c}
		u_{1x}^{*}\\
		u_{2x}^{*}\\
		u_{3x}^{*}\\
		u_{4x}^{*}\\
		u_{5x}^{*}\\
		u_{6x}^{*}\\
		u_{7x}^{*}\\
		u_{8x}^{*}\\
		u_{9x}^{*}\\
		u_{10x}^{*}\\
		u_{11x}^{*}\\
		u_{12x}^{*}\\
	\end{array}\right)=
	J_{1}P_{2,n+1}+u_{x}^{*}.\label{eq:2.8}
\end{align}
From the recurrence relations (\ref{eq:2.7}), we have
\begin{align*}
	&P_{2,n+1}=-(l_{i,j})_{12\times12}\left(
	\begin{array}{c}
		2l_{n}^{*}\\
		2k_{n}^{*}\\
		o_{n}^{*}\\
		m_{n}^{*}\\
		q_{n}^{*}\\
		p_{n}^{*}\\
		2s_{n}^{*}\\
		2r_{n}^{*}\\
		2u_{n}^{*}\\
		2t_{n}^{*}\\
		w_{n}^{*}\\
		v_{n}^{*}\\
	\end{array}\right)-\left(
	\begin{array}{c}
		2u_{2}^{*}\\
		2u_{1}^{*}\\
		u_{4}^{*}\\u_{3}^{*}\\u_{6}^{*}\\u_{5}^{*}\\2u_{8}^{*}\\2u_{7}^{*}\\2u_{10}^{*}\\2u_{9}^{*}\\u_{12}^{*}\\u_{11}^{*}
	\end{array}\right)z_{n}(t)x
	\end{align*}
	\begin{align*}
	+\left(\begin{array}{c}
		u_{2}^{*}\partial_{x}^{-1}a_{ny}^{*}+u_{2}^{*}\partial_{x}^{-1}c_{ny}^{*}+u_{6}^{*}\partial_{x}^{-1}f_{ny}^{*}+u_{8}^{*}\partial_{x}^{-1}h_{ny}^{*}+u_{10}^{*}\partial_{x}^{-1}e_{ny}^{*}+u_{12}^{*}\partial_{x}^{-1}g_{ny}^{*}-l_{ny}^{*}\\
		u_{1}\partial_{x}^{-1}a_{ny}^{*}+u_{1}\partial_{x}^{-1}c_{ny}^{*}+u_{5}\partial_{x}^{-1}g_{ny}^{*}+u_{7}\partial_{x}^{-1}j_{ny}^{*}+u_{9}\partial_{x}^{-1}d_{ny}^{*}+u_{11}^{*}\partial_{x}^{-1}f_{ny}^{*}+k_{ny}^{*}\\
		u_{4}^{*}\partial_{x}^{-1}b_{ny}^{*}+u_{8}^{*}\partial_{x}^{-1}d_{ny}^{*}+u_{10}^{*}\partial_{x}^{-1}j_{ny}^{*}-\frac{1}{2}o_{ny}^{*}\\
		u_{3}^{*}\partial_{x}^{-1}b_{ny}^{*}+u_{7}^{*}\partial_{x}^{-1}e_{ny}^{*}+u_{9}^{*}\partial_{x}^{-1}h_{ny}^{*}+\frac{1}{2}m_{ny}^{*}\\
		u_{2}^{*}\partial_{x}^{-1}g_{ny}^{*}+u_{6}^{*}\partial_{x}^{-1}a_{ny}^{*}+u_{8}^{*}\partial_{x}^{-1}e_{ny}^{*}-\frac{1}{2}q_{ny}^{*}\\
		u_{1}^{*}\partial_{x}^{-1}f_{ny}^{*}+u_{5}^{*}\partial_{x}^{-1}a_{ny}^{*}+u_{7}^{*}\partial_{x}^{-1}d_{ny}^{*}+\frac{1}{2}p_{ny}^{*}\\
		u_{2}^{*}\partial_{x}^{-1}j_{ny}^{*}+u_{4}^{*}\partial_{x}^{-1}e_{ny}^{*}+u_{6}^{*}\partial_{x}^{-1}d_{ny}^{*}+u_{8}^{*}\partial_{x}^{-1}a_{ny}^{*}+u_{8}^{*}\partial_{x}^{-1}b_{ny}^{*}+u_{10}^{*}\partial_{x}^{-1}g_{ny}^{*}-s_{ny}^{*}\\
		u_{1}^{*}\partial_{x}^{-1}h_{ny}^{*}+u_{3}^{*}\partial_{x}^{-1}d_{ny}^{*}+u_{5}^{*}\partial_{x}^{-1}e_{ny}^{*}+u_{7}^{*}\partial_{x}^{-1}a_{ny}^{*}+u_{7}^{*}\partial_{x}^{-1}b_{ny}^{*}+u_{9}^{*}\partial_{x}^{-1}f_{ny}^{*}+r_{ny}^{*}\\
		u_{2}^{*}\partial_{x}^{-1}d_{ny}^{*}+u_{4}^{*}\partial_{x}^{-1}h_{ny}^{*}+u_{8}^{*}\partial_{x}^{-1}f_{ny}^{*}+u_{10}^{*}\partial_{x}^{-1}b_{ny}^{*}+u_{10}^{*}\partial_{x}^{-1}c_{ny}^{*}+u_{12}^{*}\partial_{x}^{-1}j_{ny}^{*}-u_{ny}^{*}\\
		u_{1}^{*}\partial_{x}^{-1}e_{ny}^{*}+u_{3}^{*}\partial_{x}^{-1}j_{ny}^{*}+u_{7}^{*}\partial_{x}^{-1}g_{ny}^{*}+u_{9}^{*}\partial_{x}^{-1}b_{ny}^{*}+u_{9}^{*}\partial_{x}^{-1}c_{ny}^{*}+u_{11}^{*}\partial_{x}^{-1}h_{ny}^{*}+t_{ny}^{*}\\
		u_{2}^{*}\partial_{x}^{-1}f_{ny}^{*}+u_{10}^{*}\partial_{x}^{-1}h_{ny}^{*}+u_{12}^{*}\partial_{x}^{-1}c_{ny}^{*}-\frac{1}{2}w_{ny}^{*}\\
		u_{1}^{*}\partial_{x}^{-1}g_{ny}^{*}+u_{9}^{*}\partial_{x}^{-1}j_{ny}^{*}+u_{11}^{*}\partial_{x}^{-1}c_{ny}^{*}+\frac{1}{2}v_{ny}^{*}
	\end{array}\right)
\end{align*}
\begin{align*}
	=-L_{1}P_{2,n}-\bar{u}z_{n}(t)x+P_{2,n}^{y},
\end{align*}
where $L_{1}=(l_{i,j})_{12\times12}$ be defined as previous section.

We derive the Hamiltonian structure of (\ref{eq:2.8}) via the trace identity \cite{ref6},
\begin{align*}
	&\left\langle V_{1},\frac{\partial U_{1}}{\partial\lambda}\right\rangle = 2a^{*}+2b^{*}+2c^{*} ,\
	\left\langle V_{1},\frac{\partial U_{1}}{\partial u_{1}^{*}}\right\rangle = 2l^{*} ,\
	\left\langle V_{1},\frac{\partial U_{1}}{\partial u_{2}^{*}}\right\rangle = 2k^{*} ,\
	\left\langle V_{1},\frac{\partial U_{1}}{\partial u_{3}^{*}}\right\rangle = o^{*} ,\\
	&\left\langle V_{1},\frac{\partial U_{1}}{\partial u_{4^{*}}}\right\rangle = m^{*} ,\
	\left\langle V_{1},\frac{\partial U_{1}}{\partial u_{5}^{*}}\right\rangle = q^{*} ,\
	\left\langle V_{1},\frac{\partial U_{1}}{\partial u_{6}^{*}}\right\rangle = p^{*} ,\
	\left\langle V_{1},\frac{\partial U_{1}}{\partial u_{7}^{*}}\right\rangle = 2s^{*} ,\
	\left\langle V_{1},\frac{\partial U_{1}}{\partial u_{8}^{*}}\right\rangle = 2r^{*} ,\\
	&\left\langle V_{1},\frac{\partial U_{1}}{\partial u_{9}^{*}}\right\rangle = 2u^{*} ,\
	\left\langle V_{1},\frac{\partial U_{1}}{\partial u_{10}^{*}}\right\rangle = 2t^{*} ,\
	\left\langle V_{1},\frac{\partial U_{1}}{\partial u_{11}^{*}}\right\rangle = w^{*} ,\
	\left\langle V_{1},\frac{\partial U_{1}}{\partial u_{12}^{*}}\right\rangle = v^{*} .\
\end{align*}
Substituting the above formulate into the trace identity yields
\begin{align*}
	\frac{\delta}{\delta u}(2a^{*}+2b^{*}+2c^{*})=\lambda^{-\tau}\frac{\partial}{\partial\lambda}\lambda^{\tau}
	\left(\begin{array}{c}
		2l^{*}\\
		2k^{*}\\
		o^{*}\\
		m^{*}\\
		q^{*}\\
		p^{*}\\
		2s^{*}\\
		2r^{*}\\
		2u^{*}\\
		2t^{*}\\
		w^{*}\\
		v^{*}
	\end{array}\right),
\end{align*}
where $\tau=\frac{\lambda}{2}\frac{d}{dx}ln|tr(V_{0}^{2})|$. Balancing coefficients of each power of  in the above equality gives rise to
\begin{align*}
	\frac{\delta}{\delta u}(2a_{n+1}^{*}+2b_{n+1}^{*}+2c_{n+1}^{*})=(\tau-n)
	\left(\begin{array}{c}
		2l_{n}^{*}\\
		2k_{n}^{*}\\
		o_{n}^{*}\\
		m_{n}^{*}\\
		q_{n}^{*}\\
		p_{n}^{*}\\
		2s_{n}^{*}\\
		2r_{n}^{*}\\
		2u_{n}^{*}\\
		2t_{n}^{*}\\
		w_{n}^{*}\\
		v_{n}^{*}
	\end{array}\right).
\end{align*}
Taking $n=1$, gives $\tau=2$. Therefore we establish the following equation:
\begin{align*}
	P_{2,n+1}=
	\left(\begin{array}{c}
		2l_{n+1}^{*}\\
		2k_{n+1}^{*}\\
		o_{n+1}^{*}\\
		m_{n+1}^{*}\\
		q_{n+1}^{*}\\
		p_{n+1}^{*}\\
		2s_{n+1}^{*}\\
		2r_{n+1}^{*}\\
		2u_{n+1}^{*}\\
		2t_{n+1}^{*}\\
		w_{n+1}^{*}\\
		v_{n+1}^{*}
	\end{array}\right)=\frac{\delta}{\delta u}((\frac{-2}{n-1})(a_{n+2}^{*}+b_{n+2}^{*}+c_{n+2}^{*})).
\end{align*}
Thus, we have
\begin{align*}
u_{t}^{*}=J_{1}P_{2,n+1}=J_{1}\frac{\delta H_{n+1}^{2}}{\delta u^{*}}=-J_{1}L_{1}\frac{\delta H_{n}^{2}}{\delta u^{*}}-J_{1}\bar{u}z_{n}(t)x+J_{1}P_{2,n}^{y},\\
 H_{n+1}^{2}=(\frac{-2}{n-1})(a_{n+2}^{*}+b_{n+2}^{*}+c_{n+2}^{*}),\ n\geq0.
\end{align*}
When $n=1$, the hierarchy (\ref{eq:2.8}) reduces to the first integrable system
\begin{align*}
	u_{1t}^{*}=&\frac{1}{2}u_{1x}^{*}(\alpha(t)+\gamma(t))+\frac{1}{2}u_{5}^{*}\partial^{-1}(u_{1}^{*}u_{6}^{*}+u_{2}^{*}u_{11}^{*}+u_{8}^{*}u_{9}^{*})(\alpha(t)-\gamma(t))\\
	&+\frac{1}{2}u_{7}^{*}\partial^{-1}(u_{1}^{*}u_{8}^{*}+u_{4}^{*}u_{9}^{*}+u_{10}^{*}u_{11}^{*})(\beta(t)-\gamma(t))+\frac{1}{2}u_{9}^{*}\partial^{-1}(u_{1}^{*}u_{10}^{*}+u_{4}^{*}u_{7}^{*}+u_{5}^{*}u_{8}^{*})(\beta(t)-\alpha(t))\\
	&+\frac{1}{2}u_{11}^{*}\partial^{-1}(u_{1}^{*}u_{12}^{*}+u_{2}^{*}u_{5}^{*}+u_{7}^{*}u_{10}^{*})(\gamma(t)-\alpha(t))-2u_{1}^{*}z_{1}(t)x+u_{1x}^{*}\\
	&+(k_{1y}^{*}+u_{1}^{*}\partial_{x}^{-1}a_{1y}^{*}+u_{1}^{*}\partial_{x}^{-1}c_{1y}^{*}+u_{5}^{*}\partial_{x}^{-1}g_{1y}^{*}+u_{7}^{*}\partial_{x}^{-1}j_{1y}^{*}+u_{9}^{*}\partial_{x}^{-1}d_{1y}^{*}+u_{11}^{*}\partial_{x}^{-1}f_{1y}^{*}),\\
	u_{2t}^{*}=&\frac{1}{2}u_{2x}^{*}(\alpha(t)+\gamma(t))-\frac{1}{2}u_{6}^{*}\partial^{-1}(u_{1}^{*}u_{12}^{*}+u_{2}^{*}u_{5}^{*}+u_{7}^{*}u_{10}^{*})(\gamma(t)-\alpha(t))\\
	&-\frac{1}{2}u_{8}^{*}\partial^{-1}(u_{2}^{*}u_{7}^{*}+u_{3}^{*}u_{10}^{*}+u_{9}^{*}u_{12}^{*})(\gamma(t)-\beta(t))-\frac{1}{2}u_{10}^{*}\partial^{-1}(u_{2}^{*}u_{9}^{*}+u_{3}^{*}u_{8}^{*}+u_{6}^{*}u_{7}^{*})(\alpha(t)-\beta(t))\\
	&-\frac{1}{2}u_{12}^{*}\partial^{-1}(u_{1}^{*}u_{6}^{*}+u_{2}^{*}u_{11}^{*}+u_{8}^{*}u_{9}^{*})(\alpha(t)-\gamma(t))+2u_{2}^{*}z_{1}(t)x+u_{2x}^{*}\\
	&-(-l_{1y}^{*}+u_{2}^{*}\partial_{x}^{-1}a_{1y}^{*}+u_{2}^{*}\partial_{x}^{-1}c_{1y}^{*}+u_{6}^{*}\partial_{x}^{-1}f_{1y}^{*}+u_{8}^{*}\partial_{x}^{-1}h_{1y}^{*}+u_{10}^{*}\partial_{x}^{-1}e_{1y}^{*}+u_{12}^{*}\partial_{x}^{-1}g_{1y}^{*}),\\
	u_{3t}^{*}=&u_{3x}^{*}\beta(t)+u_{7}^{*}\partial^{-1}(u_{2}^{*}u_{9}^{*}+u_{3}^{*}u_{8}^{*}+u_{6}^{*}u_{7}^{*})(\alpha(t)-\beta(t))\\
	&+u_{9}^{*}\partial^{-1}(u_{2}^{*}u_{7}^{*}+u_{3}^{*}u_{10}^{*}+u_{9}^{*}u_{12}^{*})(\gamma(t)-\beta(t))-2u_{3}^{*}z_{1}(t)x+u_{3x}^{*}\\
	&+(m_{1y}^{*}+2u_{3}^{*}\partial_{x}^{-1}b_{1y}^{*}+2u_{7}^{*}\partial_{x}^{-1}e_{1y}^{*}+2u_{9}^{*}\partial_{x}^{-1}h_{1y}^{*}),\\
	u_{4t}^{*}=&u_{4x}^{*}\beta(t)-u_{8}^{*}\partial^{-1}(u_{1}^{*}u_{10}^{*}+u_{4}^{*}u_{7}^{*}+u_{5}^{*}u_{8}^{*})(\beta(t)-\alpha(t))\\
	&-u_{10}^{*}\partial^{-1}(u_{1}^{*}u_{8}^{*}+u_{4}^{*}u_{9}^{*}+u_{10}^{*}u_{11}^{*})(\beta(t)-\gamma(t))+2u_{4}^{*}z_{1}(t)x+u_{4x}^{*}\\
	&-(-o_{1y}^{*}+2u_{4}^{*}\partial_{x}^{-1}b_{1y}^{*}+u_{8}^{*}\partial_{x}^{-1}d_{1y}^{*}+u_{10}^{*}\partial_{x}^{-1}j_{1y}^{*}),\\
	u_{5t}^{*}=&u_{5x}^{*}\alpha(t)+u_{1}^{*}\partial^{-1}(u_{1}^{*}u_{12}^{*}+u_{2}^{*}u_{5}^{*}+u_{7}^{*}u_{10}^{*})(\gamma(t)-\alpha(t))\\
	&+u_{7}^{*}\partial^{-1}(u_{1}^{*}u_{10}^{*}+u_{4}^{*}u_{7}^{*}+u_{5}^{*}u_{8}^{*})(\beta(t)-\alpha(t))-2u_{5}^{*}z_{1}(t)x+u_{5x}^{*}\\
	&+(p_{1y}^{*}+2u_{1}^{*}\partial_{x}^{-1}f_{1y}^{*}+2u_{5}^{*}\partial_{x}^{-1}a_{1y}^{*}+2u_{7}^{*}\partial_{x}^{-1}d_{1y}^{*}),\\
	u_{6t}^{*}=&u_{6x}^{*}\alpha(t)-u_{2}^{*}\partial^{-1}(u_{1}^{*}u_{6}^{*}+u_{2}^{*}u_{11}^{*}+u_{8}^{*}u_{9}^{*})(\alpha(t)-\gamma(t))+u_{6}^{*}z_{1}(t)x\\
	&-u_{8}^{*}\partial^{-1}(u_{2}^{*}u_{9}^{*}+u_{3}^{*}u_{8}^{*}+u_{6}^{*}u_{7}^{*})(\alpha(t)-\beta(t))+2u_{6}^{8}z_{1}(t)(x)+u_{6x}^{*}\\
	&-(-q_{1y}^{*}+2u_{2}^{*}\partial_{x}^{-1}g_{1y}^{*}+2u_{6}^{*}\partial_{x}^{-1}a_{1y}^{*}+2u_{8}^{*}\partial_{x}^{-1}e_{1y}^{*}),\\
	u_{7t}^{*}=&\frac{1}{2}u_{7x}^{*}(\alpha(t)+\beta(t))+\frac{1}{2}u_{1}^{*}\partial^{-1}(u_{2}^{*}u_{7}^{*}+u_{3}^{*}u_{10}^{*}+u_{9}^{*}u_{12}^{*})(\gamma(t)-\beta(t))\\
	&+\frac{1}{2}u_{3}^{*}\partial^{-1}(u_{1}^{*}u_{10}^{*}+u_{4}^{*}u_{7}^{*}+u_{5}^{*}u_{8}^{*})(\beta(t)-\alpha(t))+\frac{1}{2}u_{5}^{*}\partial^{-1}(u_{2}^{*}u_{9}^{*}+u_{3}^{*}u_{8}^{*}+u_{6}^{*}u_{7}^{*})(\alpha(t)-\beta(t))\\
	&+\frac{1}{2}u_{9}^{*}\partial^{-1}(u_{1}^{*}u_{12}^{*}+u_{2}^{*}u_{5}^{*}+u_{7}^{*}u_{10}^{*})(\gamma(t)-\alpha(t))-2u_{7}^{*}z_{1}(t)x+u_{7x}^{*}\\
	&+(r_{1y}^{*}+u_{1}^{*}\partial_{x}^{-1}h_{1y}^{*}+u_{3}^{*}\partial_{x}^{-1}d_{1y}^{*}+u_{5}^{*}\partial_{x}^{-1}e_{1y}^{*}+u_{7}^{*}\partial_{x}^{-1}a_{1y}^{*}+u_{7}^{*}\partial_{x}^{-1}b_{1y}^{*}+u_{9}^{*}\partial_{x}^{-1}f_{1y}^{*}),\\
	u_{8t}^{*}=&\frac{1}{2}u_{8x}^{*}(\alpha(t)+\beta(t))-\frac{1}{2}u_{2}^{*}\partial^{-1}(u_{1}^{*}u_{8}^{*}+u_{4}^{*}u_{9}^{*}+u_{10}^{*}u_{11}^{*})(\beta(t)-\gamma(t))\\
	&-\frac{1}{2}u_{4}^{*}\partial^{-1}(u_{2}^{*}u_{9}^{*}+u_{3}^{*}u_{8}^{*}+u_{6}^{*}u_{7}^{*})(\alpha(t)-\beta(t))-\frac{1}{2}u_{6}^{*}\partial^{-1}(u_{1}^{*}u_{10}^{*}+u_{4}^{*}u_{7}^{*}+u_{5}^{*}u_{8}^{*})(\beta(t)-\alpha(t))\\
	&-\frac{1}{2}u_{10}^{*}\partial^{-1}(u_{1}^{*}u_{6}^{*}+u_{2}^{*}u_{11}^{*}+u_{8}^{*}u_{9}^{*})(\alpha(t)-\gamma(t))+2u_{8}^{*}z_{1}(t)x+u_{8x}^{*}\\
	&-(-s_{1y}^{*}+u_{2}^{*}\partial_{x}^{-1}j_{1y}^{*}+u_{4}^{*}\partial_{x}^{-1}e_{1y}^{*}+u_{6}^{*}\partial_{x}^{-1}d_{1y}^{*}+u_{8}^{*}\partial_{x}^{-1}a_{1y}^{*}+u_{8}^{*}\partial_{x}^{-1}b_{1y}^{*}+u_{10}^{*}\partial_{x}^{-1}g_{1y}^{*}),\\
	u_{9t}^{*}=&\frac{1}{2}u_{9x}^{*}(\beta(t)+\gamma(t))+\frac{1}{2}u_{1}^{*}\partial^{-1}(u_{2}^{*}u_{9}^{*}+u_{3}^{*}u_{8}^{*}+u_{6}^{*}u_{7}^{*})(\alpha(t)-\beta(t))\\
	&+\frac{1}{2}u_{3}^{*}\partial^{-1}(u_{1}^{*}u_{8}^{*}+u_{4}^{*}u_{9}^{*}+u_{10}^{*}u_{11}^{*})(\beta(t)-\gamma(t))+\frac{1}{2}u_{7}^{*}\partial^{-1}(u_{1}^{*}u_{6}^{*}+u_{2}^{*}u_{11}^{*}+u_{8}^{*}u_{9}^{*})(\alpha(t)-\gamma(t))\\
	&+\frac{1}{2}u_{11}^{*}\partial^{-1}(u_{2}^{*}u_{7}^{*}+u_{3}^{*}u_{10}^{*}+u_{9}^{*}u_{12}^{*})(\gamma(t)-\beta(t))-2u_{9}^{*}z_{1}(t)x+u_{9x}^{*}\\
	&+(t_{1y}^{*}+u_{1}^{*}\partial_{x}^{-1}e_{1y}^{*}+u_{3}^{*}\partial_{x}^{-1}j_{1y}^{*}+u_{7}^{*}\partial_{x}^{-1}g_{1y}^{*}+u_{9}^{*}\partial_{x}^{-1}b_{1y}^{*}+u_{9}^{*}\partial_{x}^{-1}c_{1y}^{*}+u_{11}^{*}\partial_{x}^{-1}h_{1y}^{*}),\\
	u_{10t}^{*}=&\frac{1}{2}u_{10x}^{*}(\beta(t)+\gamma(t))-\frac{1}{2}u_{2}^{*}\partial^{-1}(u_{1}^{*}u_{10}^{*}+u_{4}^{*}u_{7}^{*}+u_{5}^{*}u_{8}^{*})(\beta(t)-\alpha(t))\\
	&-\frac{1}{2}u_{4}^{*}\partial^{-1}(u_{2}^{*}u_{7}^{*}+u_{3}^{*}u_{10}^{*}+u_{9}^{*}u_{12}^{*})(\gamma(t)-\beta(t))-\frac{1}{2}u_{8}^{*}\partial^{-1}(u_{1}^{*}u_{12}^{*}+u_{2}^{*}u_{5}^{*}+u_{7}^{*}u_{10}^{*})(\gamma(t)-\alpha(t))\\
	&-\frac{1}{2}u_{12}^{*}\partial^{-1}(u_{1}^{*}u_{8}^{*}+u_{4}^{*}u_{9}^{*}+u_{10}^{*}u_{11}^{*})(\beta(t)-\gamma(t))+2u_{10}^{*}z_{1}(t)x+u_{10x}^{*}\\
	&-(-u_{1y}^{*}+u_{2}^{*}\partial_{x}^{-1}d_{1y}^{*}+u_{4}^{*}\partial_{x}^{-1}h_{1y}^{*}+u_{8}^{*}\partial_{x}^{-1}f_{1y}^{*}+u_{10}^{*}\partial_{x}^{-1}b_{1y}^{*}+u_{10}^{*}\partial_{x}^{-1}c_{1y}^{*}+u_{12}^{*}\partial_{x}^{-1}j_{1y}^{*}),\\
	u_{11t}^{*}=&u_{11x}^{*}\gamma(t)+u_{1}^{*}\partial^{-1}(u_{1}^{*}u_{6}^{*}+u_{2}^{*}u_{11}^{*}+u_{8}^{*}u_{9}^{*})(\alpha(t)-\gamma(t))\\
	&+u_{9}^{*}\partial^{-1}(u_{1}^{*}u_{8}^{*}+u_{4}^{*}u_{9}^{*}+u_{10}^{*}u_{11}^{*})(\beta(t)-\gamma(t))-2u_{11}^{*}z_{1}(t)x+u_{11x}^{*}\\
	&+(v_{1y}^{*}+2u_{1}^{*}\partial_{x}^{-1}g_{1y}^{*}+2u_{9}^{*}\partial_{x}^{-1}j_{1y}^{*}+2u_{11}^{*}\partial_{x}^{-1}c_{1y}^{*}),\\
	u_{12t}^{*}=&u_{12x}^{*}\gamma(t)-u_{2}^{*}\partial^{-1}(u_{1}^{*}u_{12}^{*}+u_{2}^{*}u_{5}^{*}+u_{7}^{*}u_{10}^{*})(\gamma(t)-\alpha(t))\\
	&-u_{10}^{*}\partial^{-1}(u_{2}^{*}u_{7}^{*}+u_{3}^{*}u_{10}^{*}+u_{9}^{*}u_{12}^{*})(\gamma(t)-\beta(t))+2u_{12}^{*}z_{1}(t)x+u_{12x}^{*}\\
	&-(-w_{1y}^{*}+2u_{2}^{*}\partial_{x}^{-1}f_{1y}^{*}+2u_{10}^{*}\partial_{x}^{-1}h_{1y}^{*}+2u_{12}^{*}\partial_{x}^{-1}c_{1y}^{*}),
\end{align*}
Therefore, we have constructed a $(2+1)$-dimensional nonisospectral integrable hierarchy on the loop algebra of the symplectic Lie algebra $\mathfrak{sp}(6)$, and present the case of $n=1$ as an example.

\section{Nonisospectral Integrable Systems on the Lie Algebra $\mathfrak{Gsp}(6)$}\label{Sec:3}
In Ref\cite{ref1},  the authors introduced a generalized Lie algebra of $\mathfrak{sp}(4)$. We employ a similar approach to consider the generalized Lie algebra of $\mathfrak{sp}(6)$, that admits a basis set as follows:
\begin{align*}
	&E_{1}=e_{11}-e_{44}, E_{2}=e_{22}-e_{55}, E_{3}=e_{33}-e_{66}, E_{4}=e_{12}-e_{54}, E_{5}=\epsilon e_{21}-\epsilon e_{45},\\
	& E_{6}=e_{13}-e_{64}, E_{7}=\epsilon e_{31}-\epsilon e_{46},E_{8}=e_{23}-e_{65}, E_{9}=\epsilon e_{32}-\epsilon e_{56}, E_{10}=\epsilon e_{16}+\epsilon e_{34},\\
	&  E_{11}=e_{43}+e_{61}, E_{12}=\epsilon e_{25}, E_{13}=e_{52},E_{14}=\epsilon e_{14},E_{15}=e_{41}, E_{16}=\epsilon e_{15}+\epsilon e_{24}, \\
	&E_{17}=e_{42}+e_{15}, E_{18}=\epsilon e_{26}+\epsilon e_{35}, E_{19}=e_{53}+e_{62}, E_{20}=\epsilon e_{36}, E_{21}=e_{63},
\end{align*}
where $\epsilon\in\mathbb{R}$. The remaining structure constants are given as follows
\begin{align*}
	&[E_{1},E_{4}]=E_{4},
	[E_{1},E_{5}]=-E_{5},
	[E_{1},E_{6}]=E_{6},
	[E_{1},E_{7}]=-E_{7},\\&
	[E_{1},E_{10}]=E_{10},
	[E_{1},E_{11}]=-E_{11},
	[E_{1},E_{14}]=2E_{14},
	[E_{1},E_{15}]=-2E_{15},\\&
	[E_{1},E_{16}]=E_{16},
	[E_{1},E_{17}]=-E_{17},
	[E_{2},E_{4}]=-E_{4},
	[E_{2},E_{5}]=E_{5},\\&
	[E_{2},E_{8}]=E_{8},
	[E_{2},E_{9}]=-E_{9},
	[E_{2},E_{12}]=2E_{12},
	[E_{2},E_{13}]=-2E_{13},\\&
	[E_{2},E_{16}]=E_{16},
	[E_{2},E_{17}]=-E_{17},
	[E_{2},E_{18}]=E_{18},
	[E_{2},E_{19}]=-E_{19},\\&
	[E_{3},E_{6}]=-E_{6},
	[E_{3},E_{7}]=E_{7},
	[E_{3},E_{8}]=-E_{8},
	[E_{3},E_{9}]=E_{9},\\&
	[E_{3},E_{10}]=E_{10},
	[E_{3},E_{11}]=-E_{11},
	[E_{3},E_{18}]=E_{18},
	[E_{3},E_{19}]=-E_{19},\\&
	[E_{3},E_{20}]=2E_{20},
	[E_{3},E_{21}]=-2E_{21},
	[E_{4},E_{5}]=\epsilon E_{1}-\epsilon E_{2},
	[E_{4},E_{7}]=-E_{9},\\&
	[E_{4},E_{8}]=E_{6},
	[E_{4},E_{11}]=-E_{19},
	[E_{4},E_{12}]=E_{16},
	[E_{4},E_{15}]=-E_{17},\\&
	[E_{4},E_{16}]=2E_{14},
	[E_{4},E_{17}]=-2E_{13},
	[E_{4},E_{18}]=E_{10},
	[E_{5},E_{6}]=\epsilon E_{8},\\&
	[E_{5},E_{9}]=-\epsilon E_{7},
	[E_{5},E_{10}]=\epsilon E_{18},
	[E_{5},E_{13}]=-\epsilon E_{17},
	[E_{5},E_{14}]=\epsilon E_{16},\\&
	[E_{5},E_{16}]=2\epsilon E_{12},
	[E_{5},E_{17}]=-2\epsilon E_{15},
	[E_{5},E_{19}]=-\epsilon E_{11},
	[E_{6},E_{7}]=\epsilon E_{1}-\epsilon E_{3},\\&
	[E_{6},E_{9}]=\epsilon E_{4},
	[E_{6},E_{10}]=2E_{14},
	[E_{6},E_{11}]=-2E_{21},
	[E_{6},E_{15}]=-E_{11},\\&
	[E_{6},E_{17}]=-E_{19},
	[E_{6},E_{18}]=E_{16},
	[E_{6},E_{20}]=E_{10},
	[E_{7},E_{8}]=-E_{5},\\&
	[E_{7},E_{10}]=2\epsilon E_{20},
	[E_{7},E_{11}]=-2\epsilon E_{15},
	[E_{7},E_{14}]=\epsilon E_{10},
	[E_{7},E_{16}]=\epsilon E_{18},\\&
	[E_{7},E_{19}]=-\epsilon E_{17},
	[E_{7},E_{21}]=-\epsilon E_{11},
	[E_{8},E_{9}]=\epsilon E_{2}-\epsilon E_{3},
	[E_{8},E_{10}]=E_{16},\\&
	[E_{8},E_{13}]=-E_{19},
	[E_{8},E_{17}]=-E_{11},
	[E_{8},E_{18}]=2E_{12},
	[E_{8},E_{19}]=-2E_{21},\\&
	[E_{8},E_{20}]=E_{18},
	[E_{9},E_{11}]=-\epsilon E_{17},
	[E_{9},E_{12}]=\epsilon E_{18},
	[E_{9},E_{16}]=\epsilon E_{10},\\&
	[E_{9},E_{18}]=2\epsilon E_{20},
	[E_{9},E_{19}]=-2\epsilon E_{13},
	[E_{9},E_{21}]=-\epsilon E_{19},
	[E_{10},E_{11}]=\epsilon E_{1}+\epsilon E_{3},\\&
	[E_{10},E_{15}]=E_{7},
	[E_{10},E_{17}]=E_{9},
	[E_{10},E_{19}]=\epsilon E_{4},
	[E_{10},E_{21}]=\epsilon E_{6},\\&
	[E_{11},E_{14}]=-\epsilon E_{6},
	[E_{11},E_{16}]=-\epsilon E_{8},
	[E_{11},E_{18}]=-E_{5},
	[E_{11},E_{20}]=-E_{7},\\&
	[E_{12},E_{13}]=\epsilon E_{2},
	[E_{12},E_{17}]=E_{5},
	[E_{12},E_{19}]=\epsilon E_{8},
	[E_{13},E_{16}]=-\epsilon E_{4},\\&
	[E_{13},E_{18}]=-E_{9},
	[E_{14},E_{15}]=\epsilon E_{1},
	[E_{14},E_{17}]=\epsilon E_{4},
	[E_{15},E_{16}]=-E_{5},\\&
	[E_{16},E_{17}]=\epsilon E_{1}+\epsilon E_{2},
	[E_{16},E_{19}]=\epsilon E_{6},
	[E_{17},E_{18}]=-E_{7},
	[E_{18},E_{19}]=\epsilon E_{2}+\epsilon E_{3},\\&
	[E_{18},E_{21}]=\epsilon E_{8},
	[E_{19},E_{20}]=-E_{9},
	[E_{20},E_{21}]=\epsilon E_{3}.
\end{align*}
We denote this generalized Lie algebra by $\mathfrak{Gsp}(6)$, and its loop algebra by $\widetilde{\mathfrak{Gsp}(6)}$. Let $U_{2},V_{2}\in\widetilde{\mathfrak{Gsp}(6)}$
\begin{align*}
	U_{2}=&E_{1}(1)+E_{2}(1)+E_{3}(1)+u_{1}'E_{10}(0)+u_{2}'E_{11}(0)+u_{3}'E_{12}(0)+u_{4}'E_{13}(0)+u_{5}'E_{14}(0)\\
	&+u_{6}'E_{15}(0)+u_{7}'E_{16}(0)+u_{8}'E_{17}(0)+u_{9}'E_{18}(0)+u_{10}'E_{19}(0)+u_{11}'E_{20}(0)+u_{12}'E_{21}(0)\\
	=&(\lambda,\lambda,\lambda,0,0,0,0,0,0,u_{1}',u_{2}',u_{3}',u_{4}',u_{5}',u_{6}',u_{7}',u_{8}',u_{9}',u_{10}',u_{11}',u_{12}')^{t},
\end{align*}
i.e.
\begin{align}
	U_{2}=\left(
	\begin{array}{cccccc}
		\lambda&0&0&\epsilon u_{5}'&\epsilon u_{7}'&\epsilon u_{1}'\\
		0&\lambda&0&\epsilon u_{7}'&\epsilon u_{3}'&\epsilon u_{9}'\\
		0&0&\lambda&\epsilon u_{1}'&\epsilon u_{9}'&\epsilon u_{11}'\\
		u_{6}'&u_{8}'&u_{2}'&-\lambda&0&0\\
		u_{8}'&u_{4}'&u_{10}'&0&-\lambda&0\\
		u_{2}'&u_{10}'&u_{12}'&0&0&-\lambda
	\end{array}\right)\label{eq:3.1}
\end{align}
and
\begin{align*}
	V_{2}=&a'E_{1}(0)+b'E_{2}(0)+c'E_{3}(0)+d'E_{4}(0)+e'E_{5}(0)+f'E_{6}(0)+g'E_{7}(0)+h'E_{8}(0)+j'E_{9}(0)\\
	&+k'E_{10}(0)+l'E_{11}(0)+m'E_{12}(0)+o'E_{13}(0)+p'E_{14}(0)+q'E_{15}(0)+r'E_{16}(0)+s'E_{17}(0)\\
	&+t'E_{18}(0)+u'E_{19}(0)+v'E_{20}(0)+w'E_{21}(0)\\
	=&(a',b',c',d',e',f',g',h',j',k',l',m',o',p',q',r',s',t',u',v',w')^{t},
\end{align*}
where $a'=\sum_{i\geq0}a_{i}\lambda^{-i},\dots,w'=\sum_{i\geq0}w_{i}\lambda^{-i},$
i.e.
\begin{align}
	V_{0}=\left(
	\begin{array}{cccccc}
		a'&d'&f'&\epsilon p'&\epsilon r'&\epsilon k'\\
		\epsilon e'&b'&h'&\epsilon r'&\epsilon m'&\epsilon t'\\
		\epsilon g'&\epsilon j'&c'&\epsilon k'&\epsilon t'&\epsilon v'\\
		q'&s'&l'&-a'&-\epsilon e'&-\epsilon g'\\
		s'&o'&u'&-d'&-b'&-\epsilon j'\\
		l'&u'&w'&-f'&-h'&-c'\\
	\end{array}\right)= \sum_{i\geq0}\left(
	\begin{array}{cccccc}
		a_{i}'&d_{i}'&f_{i}'&\epsilon p_{i}'&\epsilon r_{i}'&\epsilon k_{i}'\\
		\epsilon e_{i}'&b_{i}'&h_{i}'&\epsilon r_{i}'&\epsilon m_{i}'&\epsilon t_{i}'\\
		\epsilon g_{i}'&\epsilon j_{i}'&c_{i}'&\epsilon k_{i}'&\epsilon t_{i}'&\epsilon v_{i}'\\
		q_{i}'&s_{i}'&l_{i}'&-a_{i}'&-\epsilon e_{i}'&-\epsilon g_{i}'\\
		s_{i}'&o_{i}'&u_{i}'&-d_{i}'&-b_{i}'&-\epsilon j_{i}'\\
		l_{i}'&u_{i}'&w_{i}'&-f_{i}'&-h_{i}'&-c_{i}'\\
	\end{array}\right)\lambda^{-i}.\label{eq:3.2}
\end{align}
Consider a nonisospectral problem
\begin{align*}
	\begin{cases}
		\phi_{x}=U_{2}\phi\\
		\phi_{t}=V_{2}\phi\\
		\lambda_{t}=\sum_{i\geq0}z_{i}'(t)\lambda^{-i}
	\end{cases}
\end{align*}
 By solving stationary the zero curvature representation $V_{2x}=\frac{\partial U_{2}}{\partial\lambda}\lambda_{t}+[U_{2},V_{2}]$,
 we can obtain
 \begin{align}
 	\begin{cases}
 		a_{x}'=\epsilon u_{1}'l'-\epsilon u_{2}'k'+\epsilon u_{5}'q'-\epsilon u_{6}'p'+\epsilon u_{7}'s'-\epsilon u_{8}'r'+z'(t),\\
 		b_{x}'=\epsilon u_{3}'o'-\epsilon u_{4}'m'+\epsilon u_{7}'s'-\epsilon u_{8}'r'+\epsilon u_{9}'u'-\epsilon u_{10}'t'+z'(t),\\
 		c_{x}'=\epsilon u_{1}'l'-\epsilon u_{2}'k'+\epsilon u_{9}'u'-\epsilon u_{10}'t'+\epsilon u_{11}'w'-\epsilon u_{12}'v'+z'(t),\\
 		d_{x}'=\epsilon u_{1}'u'-\epsilon u_{4}'r'+\epsilon u_{5}'s'+\epsilon u_{7}'o'-\epsilon u_{8}'p'-\epsilon u_{10}'k',\\
 		e_{x}'=-u_{2}'t'+u_{3}'s'-u_{6}'r'+u_{7}'q'-u_{8}'m'+u_{9}'l',\\
 		f_{x}'=\epsilon u_{1}'w'-\epsilon u_{2}'p'+\epsilon u_{5}'l'+\epsilon u_{7}'u'-\epsilon u_{10}'r'-\epsilon u_{12}'k',\\
 		g_{x}'=u_{1}'q'-u_{2}'v'-u_{6}'k'-u_{8}'t'+u_{9}'s'+u_{11}'l',\\
 		h_{x}'=-\epsilon u_{2}'r'+\epsilon u_{3}'u'+\epsilon u_{7}'l'+\epsilon u_{9}'w'-\epsilon u_{10}'m'-\epsilon u_{12}'t',\\
 		j_{x}'=u_{1}'s'-u_{4}'t'-u_{8}'k'+u_{9}'o'-u_{10}'v'+u_{11}'u',\\
 		k_{x}'=2\lambda k'-u_{1}'a'-u_{1}'c'-\epsilon u_{5}'g'-\epsilon u_{7}'j'-u_{9}'d'-u_{11}'f',\\
 		l_{x}'=-2\lambda l'+u_{2}'a'+u_{2}'c'+u_{6}'f'+u_{8}'h'+\epsilon u_{10}'e'+\epsilon u_{12}'g',\\
 		m_{x}'=2\lambda m'-2u_{3}'b'-2\epsilon u_{7}'e'-2u_{9}'h',\\
 		o_{x}'=-2\lambda o'+2u_{4}'b'+2u_{8}'d'+2\epsilon u_{10}'j',\\
 		p_{x}'=2\lambda p'-2u_{1}'f'-2u_{5}'a'-2u_{7}'d',\\
 		q_{x}'=-2\lambda q'+2\epsilon u_{2}'g'+2u_{6}'a'+2\epsilon u_{8}'e',\\
 		r_{x}'=2\lambda r'-u_{1}'h'-u_{3}'d'-\epsilon u_{5}'e'-u_{7}'a'-u_{7}'b'-u_{9}'f',\\
 		s_{x}'=-2\lambda s'+\epsilon u_{2}'j'+\epsilon u_{4}'e'+u_{6}'d'+u_{8}'a'+u_{8}'b'+\epsilon u_{10}'g',\\
 		t_{x}'=2\lambda t'-\epsilon u_{1}'e'-\epsilon u_{3}'j'-\epsilon u_{7}'g'-u_{9}'b'-u_{9}'c'-u_{11}'h',\\
 		u_{x}'=-2\lambda u'+u_{2}'d'+u_{4}'h'+u_{8}'f'+u_{10}'b'+u_{10}'c'+\epsilon u_{12}'j',\\
 		v_{x}'=2\lambda v'-2\epsilon u_{1}'g'-2\epsilon u_{9}'j'-2u_{11}'c',\\
 		w_{x}'=-2\lambda w'+2u_{2}'f'+2u_{10}'h'+2u_{12}'c'.
 	\end{cases}\label{eq:3.3}
 \end{align}
 Take the initial values
 \begin{align*}
 	a_{0}'=\alpha'(t),\ b_{0}'=\beta'(t),\ c_{0}'=\gamma'(t),\ d_{0}'=e_{0}'=\dots=v_{0}'=w_{0}'=z_{0}'(t)=0.
 \end{align*}
 From (\ref{eq:3.3}), we have
 \begin{align*}
 	a_{1}'&=b_{1}'=c_{1}'=z_{1}'(t)x,\ d_{1}'=\frac{\epsilon}{2}\partial^{-1}(u_{1}'u_{10}'+u_{4}'u_{7}'+u_{5}'u_{8}')(\beta'(t)-\alpha'(t)),\\ e_{1}'&=\frac{1}{2}\partial^{-1}(u_{2}'u_{9}'+u_{3}'u_{8}'+u_{6}'u_{7}')(\alpha'(t)-\beta'(t)),\ f_{1}'=\frac{\epsilon}{2}\partial^{-1}(u_{1}'u_{12}'+u_{2}'u_{5}'+u_{7}'u_{10}')(\gamma'(t)-\alpha'(t)),\\
 	g_{1}'&=\frac{1}{2}\partial^{-1}(u_{1}'u_{6}'+u_{2}'u_{11}'+u_{8}'u_{9}')(\alpha'(t)-\gamma'(t)),\ h_{1}'=\frac{\epsilon}{2}\partial^{-1}(u_{2}'u_{7}'+u_{3}'u_{10}'+u_{9}'u_{12}')(\gamma'(t)-\beta'(t)),\\
 	j_{1}'&=\frac{1}{2}\partial^{-1}(u_{1}'u_{8}'+u_{4}'u_{9}'+u_{10}'u_{11}')(\beta'(t)-\gamma'(t)),\ k_{1}'=\frac{1}{2}u_{1}'(\alpha'(t)+\gamma'{t}),\  l_{1}'=\frac{1}{2}u_{2}'(\alpha'(t)+\gamma'{t}),\\& m_{1}'=u_{3}'\beta'(t),\ o_{1}'=u_{4}'\beta'(t),\ p_{1}'=u_{5}'\alpha'(t),\ q_{1}'=u_{6}'\alpha'(t),\ r_{1}'=\frac{1}{2}u_{7}'(\alpha'(t)+\beta'(t)),\\ s_{1}'&=\frac{1}{2}u_{8}'(\alpha'(t)+\beta'(t)),\ t_{1}'=\frac{1}{2}u_{9}'(\beta'(t)+\gamma'(t)),\ u_{1}'= \frac{1}{2}u_{10}'(\beta'(t)+\gamma'(t)),\ v_{1}'=u_{11}'\gamma'(t),\  w_{1}'=u_{12}'\gamma'(t),\\
 	k_{2}'&=\frac{1}{4}u_{1x}'(\alpha'(t)+\gamma'{t})+\frac{\epsilon}{4}u_{5}'\partial^{-1}(u_{1}'u_{6}'+u_{2}'u_{11}'+u_{8}'u_{9}')(\alpha'(t)-\gamma'(t))\\&+\frac{\epsilon}{4}u_{7}'\partial^{-1}(u_{1}'u_{8}'+u_{4}'u_{9}'+u_{10}'u_{11}')(\beta'(t)-\gamma'(t))+\frac{\epsilon}{4}u_{9}'\partial^{-1}(u_{1}'u_{10}'+u_{4}'u_{7}'+u_{5}'u_{8}')(\beta'(t)-\alpha'(t))\\&+\frac{\epsilon}{4}u_{11}'\partial^{-1}(u_{1}'u_{12}'+u_{2}'u_{5}'+u_{7}'u_{10}')(\gamma'(t)-\alpha'(t))+u_{1}'z_{1}'(t)x,\\
 	l_{2}'=&-\frac{1}{4}u_{2x}'(\alpha'(t)+\gamma'{t})+\frac{\epsilon}{4}u_{6}'\partial^{-1}(u_{1}'u_{12}'+u_{2}'u_{5}'+u_{7}'u_{10}')(\gamma'(t)-\alpha'(t))\\&+\frac{\epsilon}{4}u_{8}'\partial^{-1}(u_{2}'u_{7}'+u_{3}'u_{10}'+u_{9}'u_{12}')(\gamma'(t)-\beta'(t))+\frac{\epsilon}{4}u_{10}'\partial^{-1}(u_{2}'u_{9}'+u_{3}'u_{8}'+u_{6}'u_{7}')(\alpha'(t)-\beta'(t))\\&+\frac{\epsilon}{4}u_{12}'\partial^{-1}(u_{1}'u_{6}'+u_{2}'u_{11}'+u_{8}'u_{9}')(\alpha'(t)-\gamma'(t))+u_{2}'z_{1}'(t)x,\\
 	m_{2}'=&\frac{1}{2}u_{3x}'\beta'(t)+\frac{\epsilon}{2}u_{7}'\partial^{-1}(u_{2}'u_{9}'+u_{3}'u_{8}'+u_{6}'u_{7}')(\alpha'(t)-\beta'(t))\\&+\frac{\epsilon}{2}u_{9}'\partial^{-1}(u_{2}'u_{7}'+u_{3}'u_{10}'+u_{9}'u_{12}')(\gamma'(t)-\beta'(t))+u_{3}'z_{1}'(t)x,\\
 	o_{2}'=&-\frac{1}{2}u_{4x}'\beta'(t)+\frac{\epsilon}{2}u_{8}'\partial^{-1}(u_{1}'u_{10}'+u_{4}'u_{7}'+u_{5}'u_{8}')(\beta'(t)-\alpha'(t))\\&+\frac{\epsilon}{2}u_{10}'\partial^{-1}(u_{1}'u_{8}'+u_{4}'u_{9}'+u_{10}'u_{11}')(\beta'(t)-\gamma'(t))+u_{4}'z_{1}'(t)x,\\
 	p_{2}'=&\frac{1}{2}u_{5x}'\alpha'(t)+\frac{\epsilon}{2}u_{1}'\partial^{-1}(u_{1}'u_{12}'+u_{2}'u_{5}'+u_{7}'u_{10}')(\gamma'(t)-\alpha'(t))\\&+\frac{\epsilon}{2}u_{7}'\partial^{-1}(u_{1}'u_{10}'+u_{4}'u_{7}'+u_{5}'u_{8}')(\beta'(t)-\alpha'(t))+u_{5}'z_{1}'(t)x,\\
 	q_{2}'=&-\frac{1}{2}u_{6x}'\alpha'(t)+\frac{\epsilon}{2}u_{2}'\partial^{-1}(u_{1}'u_{6}'+u_{2}'u_{11}'+u_{8}'u_{9}')(\alpha'(t)-\gamma'(t))\\&+\frac{\epsilon}{2}u_{8}'\partial^{-1}(u_{2}'u_{9}'+u_{3}'u_{8}'+u_{6}'u_{7}')(\alpha'(t)-\beta'(t))+u_{6}'z_{1}'(t)x,\\
 	r_{2}'=&\frac{1}{4}u_{7x}'(\alpha'(t)+\beta'(t))+\frac{\epsilon}{4}u_{1}'\partial^{-1}(u_{2}'u_{7}'+u_{3}'u_{10}'+u_{9}'u_{12}')(\gamma'(t)-\beta'(t))\\&+\frac{\epsilon}{4}u_{3}'\partial^{-1}(u_{1}'u_{10}'+u_{4}'u_{7}'+u_{5}'u_{8}')(\beta'(t)-\alpha'(t))+\frac{\epsilon}{4}u_{5}'\partial^{-1}(u_{2}'u_{9}'+u_{3}'u_{8}'+u_{6}'u_{7}')(\alpha'(t)-\beta'(t))\\&+\frac{\epsilon}{4}u_{9}'\partial^{-1}(u_{1}'u_{12}'+u_{2}'u_{5}'+u_{7}'u_{10}')(\gamma'(t)-\alpha'(t))+u_{7}'z_{1}'(t)x,\\
 	s_{2}'=&-\frac{1}{4}u_{8x}'(\alpha'(t)+\beta'(t))+\frac{\epsilon}{4}u_{2}'\partial^{-1}(u_{1}'u_{8}'+u_{4}'u_{9}'+u_{10}'u_{11}')(\beta'(t)-\gamma'(t))\\&+\frac{\epsilon}{4}u_{4}'\partial^{-1}(u_{2}'u_{9}'+u_{3}'u_{8}'+u_{6}'u_{7}')(\alpha'(t)-\beta'(t))+\frac{\epsilon}{4}u_{6}'\partial^{-1}(u_{1}'u_{10}'+u_{4}'u_{7}'+u_{5}'u_{8}')(\beta'(t)-\alpha'(t))\\&+\frac{\epsilon}{4}u_{10}'\partial^{-1}(u_{1}'u_{6}'+u_{2}'u_{11}'+u_{8}'u_{9}')(\alpha'(t)-\gamma'(t))+u_{8}'z_{1}'(t)x,\\
 	t_{2}'=&\frac{1}{4}u_{9x}'(\beta'(t)+\gamma'(t))+\frac{\epsilon}{4}u_{1}'\partial^{-1}(u_{2}'u_{9}'+u_{3}'u_{8}'+u_{6}'u_{7}')(\alpha'(t)-\beta'(t))\\&+\frac{\epsilon}{4}u_{3}'\partial^{-1}(u_{1}'u_{8}'+u_{4}'u_{9}'+u_{10}'u_{11}')(\beta'(t)-\gamma'(t))+\frac{\epsilon}{4}u_{7}'\partial^{-1}(u_{1}'u_{6}'+u_{2}'u_{11}'+u_{8}'u_{9}')(\alpha'(t)-\gamma'(t))\\&+\frac{\epsilon}{4}u_{11}'\partial^{-1}(u_{2}'u_{7}'+u_{3}'u_{10}'+u_{9}'u_{12}')(\gamma'(t)-\beta'(t))+u_{9}'z_{1}'(t)x,\\
 	u_{2}'=&-\frac{1}{4}u_{10x}'(\beta'(t)+\gamma'(t))+\frac{\epsilon}{4}u_{2}'\partial^{-1}(u_{1}'u_{10}'+u_{4}'u_{7}'+u_{5}'u_{8}')(\beta'(t)-\alpha'(t))\\&+\frac{\epsilon}{4}u_{4}'\partial^{-1}(u_{2}'u_{7}'+u_{3}'u_{10}'+u_{9}'u_{12}')(\gamma'(t)-\beta'(t))+\frac{\epsilon}{4}u_{8}'\partial^{-1}(u_{1}'u_{12}'+u_{2}'u_{5}'+u_{7}'u_{10}')(\gamma'(t)-\alpha'(t))\\&+\frac{\epsilon}{4}u_{12}'\partial^{-1}(u_{1}'u_{8}'+u_{4}'u_{9}'+u_{10}'u_{11}')(\beta'(t)-\gamma'(t))+u_{10}'z_{1}'(t)x,\\
 	v_{2}'=&\frac{1}{2}u_{11x}'\gamma'(t)+\frac{\epsilon}{2}u_{1}'\partial^{-1}(u_{1}'u_{6}'+u_{2}'u_{11}'+u_{8}'u_{9}')(\alpha'(t)-\gamma'(t))\\&+\frac{\epsilon}{2}u_{9}'\partial^{-1}(u_{1}'u_{8}'+u_{4}'u_{9}'+u_{10}'u_{11}')(\beta'(t)-\gamma'(t))+u_{11}'z_{1}'(t)x,\\
 	w_{2}'=&-\frac{1}{2}u_{12x}'\gamma'(t)+\frac{\epsilon}{2}u_{2}'\partial^{-1}(u_{1}'u_{12}'+u_{2}'u_{5}'+u_{7}'u_{10}')(\gamma'(t)-\alpha'(t))\\&+\frac{\epsilon}{2}u_{10}'\partial^{-1}(u_{2}'u_{7}'+u_{3}'u_{10}'+u_{9}'u_{12}')(\gamma'(t)-\beta'(t))+u_{12}'z_{1}'(t)x\\
 	&\dots\dots
  \end{align*}
 Now, taking
 \begin{align*}
 	V_{2}^{n}=\lambda^{n}V_{2}=\sum_{i\geq0}(a_{i}',\dots,w_{i}')^{t}\lambda^{n-i},\quad V_{2,+}^{n}=\sum_{0}^{n}(a_{i}',\dots,w_{i}')^{t}\lambda^{n-i},\quad V_{2,-}^{n}=V_{2}^{n}-V_{2,+}^{n},
 \end{align*}
the zero curvature equation $\frac{\partial U_{2}}{\partial u'}u_{t}'-V_{2,+x}^{n}+\frac{\partial U_{2}}{\partial\lambda}\lambda_{t}+[U_{2},V_{2,+}^{n}]$ leads to the following Lax integrable hierarchy
\begin{align*}
	u_{t}'=\left(\begin{array}{c}
		u_{1}'\\u_{2}'\\u_{3}'\\u_{4}'\\u_{5}'\\u_{6}'\\u_{7}'\\u_{8}'\\u_{9}'\\u_{10}'\\u_{11}'\\u_{12}'
	\end{array}\right)_{t}=
\left(\begin{array}{c}
	2k_{n+1}'\\
	-2l_{n+1}'\\
	2m_{n+1}'\\
	-2o_{n+1}'\\
	2p_{n+1}'\\
	-2q_{n+1}'\\
	2r_{n+1}'\\
	-2s_{n+1}'\\
	2t_{n+1}'\\
	-2u_{n+1}'\\
	2v_{n+1}'\\
	-2w_{n+1}'\\
\end{array}\right)
\end{align*}
\begin{align}
	=\frac{1}{\epsilon}\left(\begin{array}{cccccccccccc}
		0&1&0&0&0&0&0&0&0&0&0&0\\
		-1&0&0&0&0&0&0&0&0&0&0&0\\
		0&0&0&2&0&0&0&0&0&0&0&0\\
		0&0&-2&0&0&0&0&0&0&0&0&0\\
		0&0&0&0&0&2&0&0&0&0&0&0\\
		0&0&0&0&-2&0&0&0&0&0&0&0\\
		0&0&0&0&0&0&0&1&0&0&0&0\\
		0&0&0&0&0&0&-1&0&0&0&0&0\\
		0&0&0&0&0&0&0&0&0&1&0&0\\
		0&0&0&0&0&0&0&0&-1&0&0&0\\
		0&0&0&0&0&0&0&0&0&0&0&2\\
		0&0&0&0&0&0&0&0&0&0&-2&0
	\end{array}\right)
	\left(
	\begin{array}{c}
		2\epsilon l_{n+1}'\\
		2\epsilon k_{n+1}'\\
		\epsilon o_{n+1}'\\
		\epsilon m_{n+1}'\\
		\epsilon q_{n+1}'\\
		\epsilon p_{n+1}'\\
		2\epsilon s_{n+1}'\\
		2\epsilon r_{n+1}'\\
		2\epsilon u_{n+1}'\\
		2\epsilon t_{n+1}'\\
		\epsilon w_{n+1}'\\
		\epsilon v_{n+1}'\\
	\end{array}\right)=
	J_{1}'P_{3,n+1}.\label{eq:3.4}
\end{align}
In the above expression, $J_{1}'=\frac{1}{\epsilon}J_{1}$. From the recurrence relations (\ref{eq:3.3}), we have
\begin{align*}
	P_{3,n+1}=(l'_{i,j})_{12\times12}\left(
	\begin{array}{c}
		2\epsilon l_{n}'\\
		2\epsilon k_{n}'\\
		\epsilon o_{n}'\\
		\epsilon m_{n}'\\
		\epsilon q_{n}'\\
		\epsilon p_{n}'\\
		2\epsilon s_{n}'\\
		2\epsilon r_{n}'\\
		2\epsilon u_{n}'\\
		2\epsilon t_{n}'\\
		\epsilon w_{n}'\\
		\epsilon v_{n}'\\
	\end{array}\right)+\left(
	\begin{array}{c}
		2\epsilon u_{2}'\\
		2\epsilon u_{1}'\\
		\epsilon u_{4}'\\\epsilon u_{3}'\\\epsilon u_{6}'\\\epsilon u_{5}'\\2\epsilon u_{8}'\\2\epsilon u_{7}'\\2\epsilon u_{10}'\\2\epsilon u_{9}'\\\epsilon u_{12}'\\\epsilon u_{11}'
	\end{array}\right)z_{n}'(t)x=L_{3}P_{3,n}+\bar{u}'z_{n}(t)'x.
\end{align*}
The element $l_{i,j}'$ is obtained by replacing $u_{i}$ with $u_{i}'$ in $l_{i,j}$ from $L_{1}$.
We derive the Hamiltonian structure of (\ref{eq:3.4}) via the trace identity \cite{ref6},
\begin{align*}
	&\left\langle V_{0},\frac{\partial U_{0}}{\partial\lambda}\right\rangle = 2a'+2b'+2c' ,\
	\left\langle V_{0},\frac{\partial U_{0}}{\partial u_{1}'}\right\rangle = 2\epsilon l' ,\
	\left\langle V_{0},\frac{\partial U_{0}}{\partial u_{2}'}\right\rangle = 2\epsilon k' ,\
	\left\langle V_{0},\frac{\partial U_{0}}{\partial u_{3}'}\right\rangle = \epsilon o' ,\\
	&\left\langle V_{0},\frac{\partial U_{0}}{\partial u_{4}'}\right\rangle = \epsilon m' ,\
	\left\langle V_{0},\frac{\partial U_{0}}{\partial u_{5}'}\right\rangle = \epsilon q' ,\
	\left\langle V_{0},\frac{\partial U_{0}}{\partial u_{6}'}\right\rangle = \epsilon p' ,\
	\left\langle V_{0},\frac{\partial U_{0}}{\partial u_{7}'}\right\rangle = 2\epsilon s' ,\
	\left\langle V_{0},\frac{\partial U_{0}}{\partial u_{8}'}\right\rangle = 2\epsilon r' ,\\
	&\left\langle V_{0},\frac{\partial U_{0}}{\partial u_{9}'}\right\rangle = 2\epsilon u' ,\
	\left\langle V_{0},\frac{\partial U_{0}}{\partial u_{10}'}\right\rangle = 2\epsilon t' ,\
	\left\langle V_{0},\frac{\partial U_{0}}{\partial u_{11}'}\right\rangle = \epsilon w' ,\
	\left\langle V_{0},\frac{\partial U_{0}}{\partial u_{12}'}\right\rangle = \epsilon v' .\
\end{align*}
Substituting the above formulate into the trace identity yields
\begin{align*}
	\frac{\delta}{\delta u}(2a'+2b'+2c')=\lambda^{-\tau}\frac{\partial}{\partial\lambda}\lambda^{\tau}
	\left(\begin{array}{c}
		2\epsilon l'\\
		2\epsilon k'\\
		\epsilon o'\\
		\epsilon m'\\
		\epsilon q'\\
		\epsilon p'\\
		2\epsilon s'\\
		2\epsilon r'\\
		2\epsilon u'\\
		2\epsilon t'\\
		\epsilon w'\\
		\epsilon v'
	\end{array}\right),
\end{align*}
where $\tau=\frac{\lambda}{2}\frac{d}{dx}ln|tr(V_{0}^{2})|$. Balancing coefficients of each power of  in the above equality gives rise to
\begin{align*}
	\frac{\delta}{\delta u}(2a_{n+1}'+2b_{n+1}'+2c_{n+1}')=(\tau-n)
	\left(\begin{array}{c}
		2\epsilon l_{n}'\\
		2\epsilon k_{n}'\\
		\epsilon o_{n}'\\
		\epsilon m_{n}'\\
		\epsilon q_{n}'\\
		\epsilon p_{n}'\\
		2\epsilon s_{n}'\\
		2\epsilon r_{n}'\\
		2\epsilon u_{n}'\\
		2\epsilon t_{n}'\\
		\epsilon w_{n}'\\
		\epsilon v_{n}'
	\end{array}\right).
\end{align*}
Taking $n=1$, gives $\tau=0$. Therefore we establish the following equation:
\begin{align*}
	P_{3,n+1}=
	\left(\begin{array}{c}
		2\epsilon l_{n+1}'\\
		2\epsilon k_{n+1}'\\
		\epsilon o_{n+1}'\\
		\epsilon m_{n+1}'\\
		\epsilon q_{n+1}'\\
		\epsilon p_{n+1}'\\
		2\epsilon s_{n+1}'\\
		2\epsilon r_{n+1}'\\
		2\epsilon u_{n+1}'\\
		2\epsilon t_{n+1}'\\
		\epsilon w_{n+1}'\\
		\epsilon v_{n+1}'
	\end{array}\right)=\frac{\delta}{\delta u}((\frac{-2}{n+1})(a_{n+2}'+b_{n+2}'+c_{n+2}')).
\end{align*}
Thus, we get
\begin{align*}
	u_{t}'=J_{1}'P_{3,n+1}=J_{1}'\frac{\delta H_{n+1}^{3}}{\delta u'}=J_{1}'L_{1}'\frac{\delta H_{n}^{3}}{\delta u'}+J_{1}'\bar{u}'z_{n}(t)x,\\ H_{n+1}^{1}=(\frac{-2}{n+1})(a_{n+2}'+b_{n+2}'+c_{n+2}'),\ n\geq0.
\end{align*}
When $n=1$, the hierarchy (\ref{eq:3.4}) reduces to the first integrable system
\begin{align*}
u_{1t}'=&\frac{1}{2}u_{1x}'(\alpha'(t)+\gamma'{t})
+\frac{\epsilon}{2}u_{5}'\partial^{-1}(u_{1}'u_{6}'+u_{2}'u_{11}'+u_{8}'u_{9}')(\alpha'(t)-\gamma'(t))\\
&+\frac{\epsilon}{2}u_{7}'\partial^{-1}(u_{1}'u_{8}'+u_{4}'u_{9}'+u_{10}'u_{11}')(\beta'(t)-\gamma'(t))\\
&+\frac{\epsilon}{2}u_{9}'\partial^{-1}(u_{1}'u_{10}'+u_{4}'u_{7}'+u_{5}'u_{8}')(\beta'(t)-\alpha'(t))\\
&+\frac{\epsilon}{2}u_{11}'\partial^{-1}(u_{1}'u_{12}'+u_{2}'u_{5}'+u_{7}'u_{10}')(\gamma'(t)-\alpha'(t))+2u_{1}'z_{1}'(t)x,\\
u_{2t}'=&\frac{1}{2}u_{2x}'(\alpha'(t)+\gamma'{t})-\frac{\epsilon}{2}u_{6}'\partial^{-1}(u_{1}'u_{12}'+u_{2}'u_{5}'+u_{7}'u_{10}')(\gamma'(t)-\alpha'(t))\\
&-\frac{\epsilon}{2}u_{8}'\partial^{-1}(u_{2}'u_{7}'+u_{3}'u_{10}'+u_{9}'u_{12}')(\gamma'(t)-\beta'(t))\\
&-\frac{\epsilon}{2}u_{10}'\partial^{-1}(u_{2}'u_{9}'+u_{3}'u_{8}'+u_{6}'u_{7}')(\alpha'(t)-\beta'(t))\\
&-\frac{\epsilon}{2}u_{12}'\partial^{-1}(u_{1}'u_{6}'+u_{2}'u_{11}'+u_{8}'u_{9}')(\alpha'(t)-\gamma'(t))-2u_{2}'z_{1}'(t)x,\\
	u_{3t}'=&u_{3x}'\beta'(t)+\epsilon u_{7}'\partial^{-1}(u_{2}'u_{9}'+u_{3}'u_{8}'+u_{6}'u_{7}')(\alpha'(t)-\beta'(t))\\&+\epsilon u_{9}'\partial^{-1}(u_{2}'u_{7}'+u_{3}'u_{10}'+u_{9}'u_{12}')(\gamma'(t)-\beta'(t))+2u_{3}'z_{1}'(t)x,\\
	u_{4t}'=&u_{4x}'\beta'(t)-\epsilon u_{8}'\partial^{-1}(u_{1}'u_{10}'+u_{4}'u_{7}'+u_{5}'u_{8}')(\beta'(t)-\alpha'(t))\\&-\epsilon u_{10}'\partial^{-1}(u_{1}'u_{8}'+u_{4}'u_{9}'+u_{10}'u_{11}')(\beta'(t)-\gamma'(t))-2u_{4}'z_{1}'(t)x,\\
	u_{5t}'=&u_{5x}'\alpha'(t)+\epsilon u_{1}'\partial^{-1}(u_{1}'u_{12}'+u_{2}'u_{5}'+u_{7}'u_{10}')(\gamma'(t)-\alpha'(t))\\&+\epsilon u_{7}'\partial^{-1}(u_{1}'u_{10}'+u_{4}'u_{7}'+u_{5}'u_{8}')(\beta'(t)-\alpha'(t))+2u_{5}'z_{1}'(t)x,\\
	u_{6t}'=&u_{6x}'\alpha'(t)-\epsilon u_{2}'\partial^{-1}(u_{1}'u_{6}'+u_{2}'u_{11}'+u_{8}'u_{9}')(\alpha'(t)-\gamma'(t))\\&-\epsilon u_{8}'\partial^{-1}(u_{2}'u_{9}'+u_{3}'u_{8}'+u_{6}'u_{7}')(\alpha'(t)-\beta'(t))-2u_{6}'z_{1}'(t)x,\\
	u_{7t}'=&\frac{1}{2}u_{7x}'(\alpha'(t)+\beta'(t))+\frac{\epsilon}{2}u_{1}'\partial^{-1}(u_{2}'u_{7}'+u_{3}'u_{10}'+u_{9}'u_{12}')(\gamma'(t)-\beta'(t))\\&+\frac{\epsilon}{2}u_{3}'\partial^{-1}(u_{1}'u_{10}'+u_{4}'u_{7}'+u_{5}'u_{8}')(\beta'(t)-\alpha'(t))+\frac{\epsilon}{2}u_{5}'\partial^{-1}(u_{2}'u_{9}'+u_{3}'u_{8}'+u_{6}'u_{7}')(\alpha'(t)-\beta'(t))\\&+\frac{\epsilon}{2}u_{9}'\partial^{-1}(u_{1}'u_{12}'+u_{2}'u_{5}'+u_{7}'u_{10}')(\gamma'(t)-\alpha'(t))+2u_{7}'z_{1}'(t)x,\\
	u_{8t}'=&\frac{1}{2}u_{8x}'(\alpha'(t)+\beta'(t))-\frac{\epsilon}{2}u_{2}'\partial^{-1}(u_{1}'u_{8}'+u_{4}'u_{9}'+u_{10}'u_{11}')(\beta'(t)-\gamma'(t))\\&-\frac{\epsilon}{2}u_{4}'\partial^{-1}(u_{2}'u_{9}'+u_{3}'u_{8}'+u_{6}'u_{7}')(\alpha'(t)-\beta'(t))-\frac{\epsilon}{2}u_{6}'\partial^{-1}(u_{1}'u_{10}'+u_{4}'u_{7}'+u_{5}'u_{8}')(\beta'(t)-\alpha'(t))\\&-\frac{\epsilon}{2}u_{10}'\partial^{-1}(u_{1}'u_{6}'+u_{2}'u_{11}'+u_{8}'u_{9}')(\alpha'(t)-\gamma'(t))-2u_{8}'z_{1}'(t)x,\\
	u_{9t}'=&\frac{1}{2}u_{9x}'(\beta'(t)+\gamma'(t))+\frac{\epsilon}{2}u_{1}'\partial^{-1}(u_{2}'u_{9}'+u_{3}'u_{8}'+u_{6}'u_{7}')(\alpha'(t)-\beta'(t))\\&+\frac{\epsilon}{2}u_{3}'\partial^{-1}(u_{1}'u_{8}'+u_{4}'u_{9}'+u_{10}'u_{11}')(\beta'(t)-\gamma'(t))+\frac{\epsilon}{2}u_{7}'\partial^{-1}(u_{1}'u_{6}'+u_{2}'u_{11}'+u_{8}'u_{9}')(\alpha'(t)-\gamma'(t))\\&+\frac{\epsilon}{2}u_{11}'\partial^{-1}(u_{2}'u_{7}'+u_{3}'u_{10}'+u_{9}'u_{12}')(\gamma'(t)-\beta'(t))+2u_{9}'z_{1}'(t)x,\\
	u_{10t}'=&\frac{1}{2}u_{10x}'(\beta'(t)+\gamma'(t))-\frac{\epsilon}{2}u_{2}'\partial^{-1}(u_{1}'u_{10}'+u_{4}'u_{7}'+u_{5}'u_{8}')(\beta'(t)-\alpha'(t))\\&-\frac{\epsilon}{2}u_{4}'\partial^{-1}(u_{2}'u_{7}'+u_{3}'u_{10}'+u_{9}'u_{12}')(\gamma'(t)-\beta'(t))-\frac{\epsilon}{2}u_{8}'\partial^{-1}(u_{1}'u_{12}'+u_{2}'u_{5}'+u_{7}'u_{10}')(\gamma'(t)-\alpha'(t))\\&-\frac{\epsilon}{2}u_{12}'\partial^{-1}(u_{1}'u_{8}'+u_{4}'u_{9}'+u_{10}'u_{11}')(\beta'(t)-\gamma'(t))-2u_{10}'z_{1}'(t)x,\\
	u_{11t}'=&u_{11x}'\gamma'(t)+\epsilon u_{1}'\partial^{-1}(u_{1}'u_{6}'+u_{2}'u_{11}'+u_{8}'u_{9}')(\alpha'(t)-\gamma'(t))\\&+\epsilon u_{9}'\partial^{-1}(u_{1}'u_{8}'+u_{4}'u_{9}'+u_{10}'u_{11}')(\beta'(t)-\gamma'(t))+2u_{11}'z_{1}'(t)x,\\
	u_{12t}'=&u_{12x}'\gamma'(t)-\epsilon u_{2}'\partial^{-1}(u_{1}'u_{12}'+u_{2}'u_{5}'+u_{7}'u_{10}')(\gamma'(t)-\alpha'(t))\\&-\epsilon u_{10}'\partial^{-1}(u_{2}'u_{7}'+u_{3}'u_{10}'+u_{9}'u_{12}')(\gamma'(t)-\beta'(t))-2u_{12}'z_{1}'(t)x.
\end{align*}
Thus we construct a nonisospectral integrable hierarchy on the loop algebra of the generalized symplectic Lie algebra $\mathfrak{Gsp}(6)$, with the $n=1$ case presented as a specific example.
\section*{Acknowledgment}
This research was supported by the National Natural Science Foundation of China (No. 11961049, 10601219) and by the Key Project of Jiangxi Natural Science Foundation grant (No. 20232ACB201004).

	\end{document}